\documentclass[paper=a4, fontsize=11pt]{scrartcl} % A4 paper and 11pt font size

\usepackage[T1]{fontenc} % Use 8-bit encoding that has 256 glyphs
\usepackage{fourier} % Use the Adobe Utopia font for the document - comment this line to return to the LaTeX default
\usepackage[english]{babel} % English language/hyphenation
\usepackage{amsmath,amsfonts,amsthm} % Math packages
\usepackage{epsfig}
\usepackage{multirow}
\usepackage{amssymb}
\usepackage{tabularx}
\usepackage{tikz}
\usepackage{arydshln}
\usepackage{setspace}
\usepackage{soul}

\usepackage{lipsum} % Used for inserting dummy 'Lorem ipsum' text into the template

\usepackage{sectsty} % Allows customizing section commands
\allsectionsfont{\centering \normalfont\scshape} % Make all sections centered, the default font and small caps

\usepackage{fancyhdr} % Custom headers and footers
\usetikzlibrary{shapes,snakes}
\usetikzlibrary{automata, positioning, arrows}

\numberwithin{equation}{section} % Number equations within sections (i.e. 1.1, 1.2, 2.1, 2.2 instead of 1, 2, 3, 4)
\numberwithin{figure}{section} % Number figures within sections (i.e. 1.1, 1.2, 2.1, 2.2 instead of 1, 2, 3, 4)
\numberwithin{table}{section} % Number tables within sections (i.e. 1.1, 1.2, 2.1, 2.2 instead of 1, 2, 3, 4)

\newcommand{\horrule}[1]{\rule{\linewidth}{#1}} % Create horizontal rule command with 1 argument of height

\title{	
\normalfont \normalsize
\textsc{Notes on Reliability Theory} \\ [25pt] % Your university, school and/or department name(s)
\horrule{0.5pt} \\[0.4cm] % Thin top horizontal rule
\huge Durability and Availability of Erasure-Coded Storage Systems with Concurrent Maintenance  \\ % The assignment title
\horrule{2pt} \\[0.5cm] % Thick bottom horizontal rule
}

\author{Suayb S. Arslan} % Your name

\date{\footnotesize{``\emph{All truths are easy to understand once they are discovered; the point is to discover them.}" G. Galilei. \\
\normalsize{Version 2.0 -- January 2023} }}
\begin{document}

\maketitle % Print the title

%----------------------------------------------------------------------------------------
%	PROBLEM 1
%----------------------------------------------------------------------------------------

\section*{Preface}

\footnotesize{
\hspace{5mm} This initial version of this document was written back in 2014 for the sole purpose of providing fundamentals of reliability theory as well as to identify the theoretical types of machinery for the prediction of durability/availability of erasure-coded storage systems. Since the definition of a "system" is too broad, we specifically focus on warm and cold storage systems where the data is stored in a distributed fashion across different storage units with or without continuous (full duty-cycle) operation. 

\hspace{5mm}  The contents of this document are dedicated to a review of fundamentals, a few major improved stochastic models, and several contributions of my work relevant to the field. One of the interesting contributions of this document is the introduction of the most general form of Markov models for the estimation of mean time to failure numbers. This work was partially later published in \textit{IEEE Transactions on Reliability}. Very good approximations for the closed-form solutions for this general model are also investigated. Various storage configurations under different policies are compared using such advanced models. Later in a subsequent chapter, we have also considered multi-dimensional Markov models to address detached drive-medium combinations such as those found in optical disk and tape storage systems. It is not hard to anticipate such a system structure would most likely be part of future DNA storage libraries and hence find a plethora of interesting applications. This work is partially published in \textit{Elsevier Reliability and System Safety}.  

\hspace{5mm} Topics that include simulation modelings for more accurate estimations are included towards the end of the document by noting the deficiencies of the simplified canonical as well as more complex Markov models, due mainly to the stationary and static nature of Markovinity. Throughout the document, we shall focus on concurrently maintained systems although the discussions will only slightly change for the systems repaired one device at a time. The document is still under construction and future versions might likely include newer models and novel approaches to enrich the present contents. Some background on probability and coding theory might be expected that are briefly mentioned in the beginning of the document.  \\
}

\textit{January 14th, 2023 \\ Boston, MA, USA.} 

\normalsize
\newpage
\tableofcontents

\newpage
\section{Device Reliability Basics}

\hspace{5mm}  When a brand new product is put into service, it performs functional operations satisfactorily for a  period of
time, called \emph{useful time} period, before eventually a failure occurs and the device becomes no longer able to
respond to incoming user requests. For a device component, the observed time to failure ($TTF$) is a
continuous random variable with a probability density function
$f_{TTF}(t)$, representing the lifetime of the product until the first permanent failure. The
probability of failure of the device can be found using the
cumulative distribution function (CDF) of $TTF$  as follows,
\begin{eqnarray}
F_{TTF}(t) \triangleq Pr\{TTF \leq t\} = \int_{0}^{t} f_{TTF}(y) dy, \ \ t
> 0
\end{eqnarray}

\hspace{5mm}  We can think of $F_{TTF}(t)$ as an \emph{unreliability} measure between
time 0 and $x$. On the other hand, the reliability function $R(t)$
is defined as,
\begin{align}
R(t) \triangleq 1 - F_{TTF}(t)
= \int_{t}^{\infty} f_{TTF}(y) dy \label{Rel}
\end{align}

\begin{figure}[b!]
\centering
\includegraphics[angle=0, height=60mm, width=115mm]{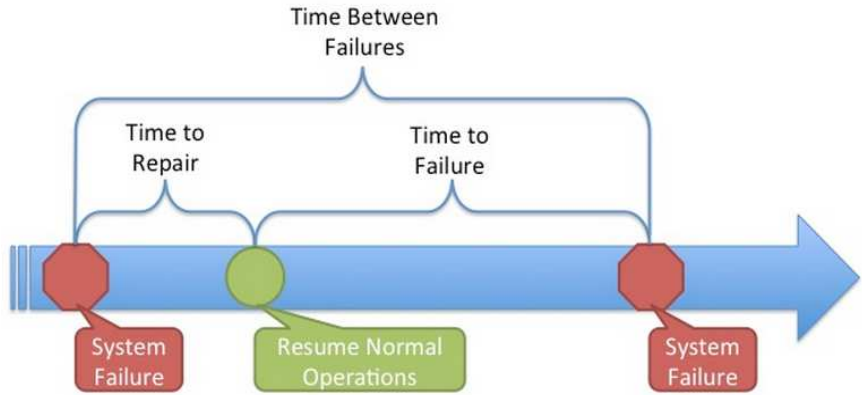}
\caption{Mean time to failure and repair visualized in time domain.}\label{fig:MTTFMTTR}
\end{figure}

\hspace{5mm}  In other words, reliability is the probability of having no failures before
time $t$ and is related to CDF of $TFF$ through  Eqn. (\ref{Rel}).
Note that Eqn. (\ref{Rel}) implies that we have
$f_{TTF}(t)=-dR(t)/dx$. In a real world, it may not be possible to estimate the
distribution function of $TFF$ directly from the available physical
information. A useful function for clarifying the relationship
between physical modes of failure and the probability distribution
of $TFF$ is known as the \emph{hazard rate} function or
\emph{failure rate} function, frequently denoted as $h_{TTF}(t)$. This function
is defined as follows,
\begin{eqnarray}
h_{TTF}(t) \triangleq \frac{f_{TTF}(t)}{R(t)}  = -\frac{dR(t)}{R(t)dt}  \label{ODE}
\end{eqnarray}

\hspace{5mm}  Solution of the first order ordinary differential equation
(\ref{ODE}) yields the relationship 
\begin{align}
    h_{TTF}(t) =-d(\ln(R(t)))/dt
\end{align}
with the initial condition $R(0)=1$. Note that  knowing the hazard rate is
equivalent to knowing the  distribution. Thus, when we talk about reliability of a system, we interchangeably use these functions to quantify it.  Mean time expected until the first failure of a piece of equipment or a total data loss is one of the most popular measures of reliability.  Mean time to failure (MTTF) is defined to be the expected value of the random
variable $TTF$ and is given by
\begin{eqnarray}
MTTF &=& \mathbb{E}[TTF]
= \int_{0}^{\infty} R(t) dt \Leftrightarrow \lim_{t \rightarrow \infty} t R(t) = 0 \label{IFF}
\end{eqnarray}
where $\mathbb{E}[.]$ is the expectation operator. Note that Eqn. (\ref{IFF}) is true for distributions whose mean exists.
 For the rest of our discussions,  the subscript $TTF$ is dropped for notation simplicity.

\hspace{5mm}  Annualized failure rate (AFR) for a device is frequently used to estimate the failure probability of a
device or a component after a full-time year use. In the
conventional approach, time between the start of the operation and the point when failure happens are assumed to be independent and
exponentially distributed with a constant rate $\lambda$. Therefore AFR is given by $AFR = 1 - R(t) = 1 - e^{-\lambda t}$,
where $\lambda = 1/MTTF$ and $x$ is the running
time index in hours. MTTF is  reported in hours and since there are
8760 hours in a year, $AFR = 1 - e^{-8760/MTTF}$.  Since $8760/MTTF \ll 1$, then
 $AFR \approx  8760 / MTTF$. Alternatively, if MTTF is expressed in years, then  $AFR \approx  1 / MTTF$.

\subsection{Disks in Real Life and Relevant Research}

 \hspace{5mm}  In this document, if a device fails we assume that there is an external agent who can repair it. The time needed for the agent to replace or repair a failed device is also a random variable whose mean is called Mean Time To Repair (MTTR). To avoid MTTR, many companies purchase spare products/hot swaps so that a replacement can be installed quickly. Generally, however, customers inquire about the turn-around time of repairing a product, and indirectly, this would eventually fall into the MTTR category. Finally, Mean Time Between Failure (MTBF) is a reliability measure used to give the time between two consecutive failures of the same device or system component \cite{hoyland1994}. This is the most common inquiry about a product's life span, and is important in the decision-making process of the end user. Fig. \ref{fig:MTTFMTTR} summarizes these terms using a timeline of an operational storage system.

 \begin{figure}[t!]
\centering
\includegraphics[angle=0, height=67mm, width=92mm]{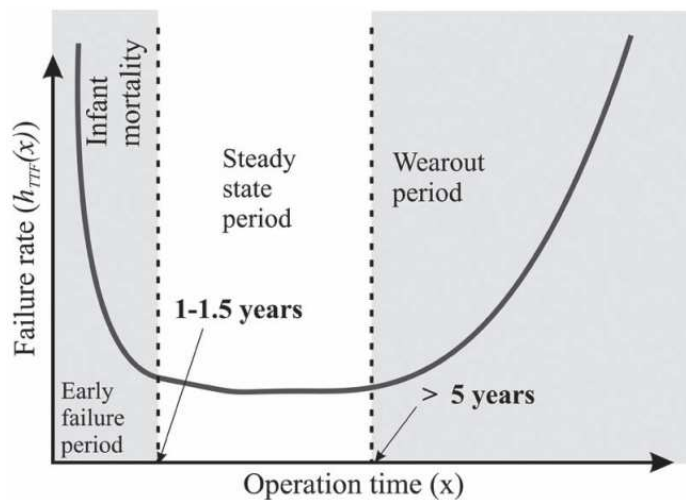}
\caption{Hazard rate pattern for hard disk drives as a function of operation
time \cite{Suayb2}.}\label{fig:Bathtub}
\end{figure}

 \hspace{5mm}  It is shown in various research articles that average replacement rate of component disks is around twenty times much greater than are the theoretically predicted MTTF values i.e., predicted MTTF values are observed to be an underestimator \cite{schroeder2007}. It is demonstrated that disk failure rates show a
“bathtub” curve as shown in Fig. \ref{fig:Bathtub}. Additionally, contrary to conventional homogenous stochastic models, hard disk replacement rates do not enter into steady state. After few years of
use, drives (majority of which are disks) are observed to enter into wear-out period in which the failure rates steadily increase over time. Time between failures are shown to give much better
fit with Weibull or gamma distributions instead of widely used exponential distribution \cite{schroeder2007}, \cite{yang1999}. There is a considerable amount of evidence that disk failures that are placed in the same batch show significant correlations, which is hard to quantify in a
number of applications \cite{pinheiro2007}.

 \hspace{5mm} There have been many efforts in industry as well as in academia for accurately predicting the reliability of large-scale storage systems in terms of mean lifetime to failure rates. For example, an accurate yet complicated model is developed to include
catastrophic failures and usage dependent data corruptions in \cite{elerath2007}. The authors specifically pointed out that component failure rates have little, if not any, to share with the failure rate of the whole storage system. The times between successive
system failures are reported to be relatively larger than what conventional models suggest, even though each component disk hazard rate is increasing \cite{ascher1999}. Disk scrubbing is introduced and used in \cite{schwarz2004} as a remedy for latent defects that are usually
independent of the size, use and the operation of disks. The latter study also uses homogenous Poisson model for reliability estimations. The heavy-tailedness of disk failure lifetime is quite well known and modeled in literature \cite{arslanzeydan2019}. One of the most critical trend is to use collected data (For instance, Backblaze provides such datasets for research \cite{backblaze0}) to model data storage system reliability for predictive maintenance \cite{zeyars2020}. Later in the document (Section 6) we will demonstrate one specific use case of data for modeling complex systems to help with the conventional mathematical tools for estimating dependibility of storage systems.

 \section{Coding Basics for Reliability}

 \hspace{5mm} In practical data storage systems, the information contained in data nodes or devices are encoded to generate some form of redundancy in order to make the user data robust against device failures/defects or random/burst errors. Whenever storage nodes/devices fail, the system controller identifies those device failures and hence the failure locations are easily  determined. From a coding perspective, these failures are regarded as erasures. Therefore, the process of creating this redundancy is called erasure correction coding or simply \emph{erasure coding}. Erasure coding is used to increase the reliability of the overall storage system using systematically generated redundant data. Error/erasure correction coding is quite old and inclusive  subject, originated almost fifty years ago. We will only consider erasure coding tailored to data storage systems in the rest of this document.

  \begin{figure}[t!]
\centering
\includegraphics[angle=0, height=12mm, width=85mm]{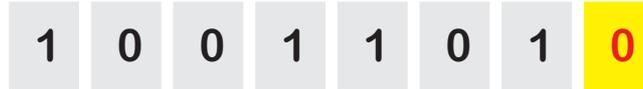}
\caption{A simple parity coding that appends a parity bit at the end of the block in order to keep the total number of 1's even. }\label{fig:PCE}
\end{figure}

 \hspace{5mm} Let us provide a short description to help motivate the rest of the content. Suppose that we have $n$ devices that contain $m$ blocks of  user data. Then, we use erasure coding to generate redundancy that consists of a subdivision of that redundancy into $c$ blocks and storing them in $c$ different devices i.e., the total number of devices is $n = m +c$. A space optimal (also known as Maximum Distance Separable (MDS)) coding scheme allows us to reconstruct the $m$ device information from the information contained in any $m$ out of $n$ devices. Thus one can realize that MDS codes enable a regular and understandable recovery mechanism. A simple instance of MDS codes is the XOR-based \emph{parity coding} scheme illustrated in Fig. \ref{fig:PCE}. Using the ``even" convention, the total number of binary 1's are kept even by adding one more binary digit at the end of the block. In linear algebra terms, for a given $(n,m)$ linear code defined by the generator matrix $\textbf{G}$, if all $m \times m$ submatrices of $\textbf{G}$ are invertible, then this linear code can tolerate any combination of $c$ erasures. The class of codes that has this property are called MDS codes.

 \hspace{5mm} We can divide data storage-specific erasure codes into three classes : flat MDS codes, parity--check array
codes and flat XOR--based codes. Reed-Solomon codes \cite{RScodes1960} are example of flat MDS codes while exhibiting computationally demanding implementation in practice.  Most array codes are space--optimal in the number of devices and less computationally expensive than Reed-Solomon, but are typically only two or
three device fault--tolerant \cite{Blaum}, \cite{Tomislav}, \cite{Star}. Since the main purpose of this document is to present the reliability issues of these codes with respect to maintainable systems, the details of these codes are not presented. A good tutorial about erasure coding for storage applications can be found at
\cite{Plank}. 

 \hspace{5mm} On the other hand, most of the modern codes are flat XOR-based codes and used to protect the data in storage applications. They are based on low complexity, low locality\footnote{In the context of erasure coding for storage, the term locality means the number of storage units or coding symbols that need to be accessed for recovering a specific unit or symbol. Locality can be considered in terms of purely coding perspective or coding+allocation strategy perspective. The context in which the locality is referred matters and may bring out different conclusions.}, irregular erasure correction coding i.e., codes with irregular failure tolerance, namely Low Density Parity Check (LDPC) codes  \cite{gallager1963} and fountain (LT) codes \cite{Luby2002}. Later implementations of these codes have taken into account the repairability and update constraints and provided amenable structures to help with the system maintenance \cite{founsure2021}.   Particularly, due to their irregular structure, they are shown to achieve good performance in the repair bandwidth-storage trade-off \cite{dimakis2010}, \cite{pourmandi2022}. Such codes, typically constructed using graph theory or concatenation of multiple algebraic constructions (Product codes) \cite{cideciyan2017}, are designed by considering the overall storage problem as a whole and they necessitate major changes to modeling assumptions which lead to interesting open problems regarding the reliability/ durability/ availability of the storage systems. In addition, choosing the appropriate code that address the data durability constraints do not necessarily solve the overall reliability problem, as the allocation of resources \cite{shi2013}, \cite{narman2018} and coded data on the available hardware and networks is shown to be profoundly important and relevant \cite{Arslan2023}. These and the like open problems shall be articulated and covered in the later chapters of the document.

\section{A Markov Reliability Model with Concurrent Maintenance}

\hspace{5mm} Traditionally, Markov models are used to evaluate the reliability of erasure-coded storage systems. Since we deal with system failures or total data loss scenarios, the class of Markov models we consider has absorbing states. An absorbing state is a system failure state or a total data loss state that, once entered, cannot be departed.
Most models assume a RAID-like setting (i.e. MDS code -- regular fault tolerance as explained previous sections), independence between failures and exponentially distributed failure/rebuild times. For this case, Markov models are perfect fit for modeling the system behavior. In fact, Markov models have provided a great deal of insight into the sensitivity of disk failure and repair on system reliability, despite these models generally capture an extremely simplistic view of an actual system.

\begin{figure}[t!]
\centering
\includegraphics[angle=0, height=30mm, width=70mm]{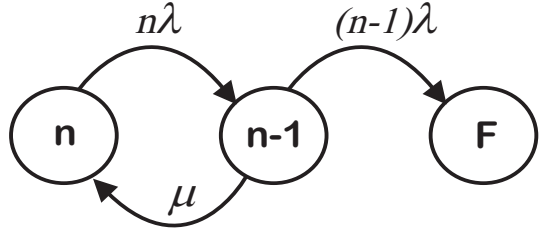}
\caption{A canonical continuous time Markov model for the reliability estimation of parity-coded data.}\label{fig:MM4}
\end{figure}

\hspace{5mm} The canonical Markov model used in storage systems is based on parity coding, which can tolerate only one device failure out of say, $n$ storage units or devices. There are a total of three states; state $n$ represents the state of all devices are active and operational, state $n-1$ represents the state of one failed device and state $F$ is the state of failure because at least two devices are failed and data integrity is lost. It is assumed that all devices fail or are repaired at the constant rates $\lambda$ and $\mu$, respectively. State diagram of the Markov model is shown in Fig. \ref{fig:MM4}. Let us start with the following lemma and use its result to talk about holding times i.e., the time it takes to hold in a specific state.

\hspace{5mm} \textbf{Lemma 1:} \emph{Let $X_1,\dots, X_n$ be independent exponentially distributed random variables with rates $\lambda_i$ for $i=1,\dots,n$. Then, the distribution function of $\min\{X_1,\dots,X_n\}$ is exponential with rate $\sum_i \lambda_i$ and the probability that the minimum is $X_s$, $s\in \{1,\dots,n\}$ is given by $\lambda_s/\sum_i\lambda_i$}.

\hspace{5mm} PROOF: Let us consider the probability,
\begin{eqnarray}
Pr\{\min\{X_1,\dots,X_n\} > t\} = \prod_i Pr\{X_i > t\} = e^{-(\sum_i\lambda_i)t}
\end{eqnarray}
which means that the cumulative distribution function of $\min\{X_1,\dots,X_n\}$ is that of an exponential distribution with rate $\sum_i\lambda_i$. One can realize that it is not the case with $\max\{X_1,\dots,X_n\}$. Furthermore,
\begin{eqnarray}
Pr\{X_s\textrm{ is the minimum}\} &=& Pr\{X_s < X_j \textrm{ for } j \not= s\} \\
&=& \int_{0}^{\infty} Pr\left\{X_s < X_j \textrm{ for } j \not= s | X_s = t\right\} \lambda_se^{-\lambda_st}dt \\
&=& \int_{0}^{\infty} \lambda_s e^{-\lambda_st} \prod_{j \not= s} e^{-\lambda_jt} dt = \lambda_s  \int_{0}^{\infty} e^{-(\sum_j\lambda_j)t} \\
&=& \lambda_s / (\lambda_1 + \dots + \lambda_n)
\end{eqnarray}
as desired.  \hfill $\square$

\hspace{5mm} Since in each state, the label represents the number of devices failing with the same rate $\lambda$, the transition out of state $n$ happens with rate $n\lambda$. Using the result of lemma 1, we can conclude that the holding time $Y_n$ is exponentially distributed with rate $n\lambda$. Similarly, the holding time in state $n-1$, $Y_{n-1}$ is exponentially distributed with rate $(n-1)\lambda + \mu$. The probability that the system transitions from state $n-1$ to state $F$ (failed state)  with probability $p_{F} = (n-1)\lambda/((n-1)\lambda + \mu)$.

\hspace{5mm} Traditional storage system reliability analysis uses a metric called the mean
time to data loss (MTTDL). MTTDL is the expected (average) time to enter state $F$. MTTDL is not a direct measure of reliability but is an average based on the reliability, given by  $\int_0^{\infty} R(t)dt$. MTTDL might be very hard to compute for most of the stochastic models that describe the real life failure phenomenon.  Fortunately, there are explicit mathematical methods to compute the MTTDL for simpler Markov models such as given in Fig. \ref{fig:MM4}.

\hspace{5mm} Let us assume we visit state $n-1$ repeatedly $K$ many times before the whole system ends up in state $F$. Thus the total time before data loss is given by
\begin{eqnarray}
Y_{DL} = \sum_{i=1}^K (Y_n^{(i)} + Y_{n-1}^{(i)})
\end{eqnarray}
where $Y_n^{(i)}$ is the $i$-th holding time. By definition, MTTDL is $\mathbb{E}[Y_{DL}]$. Conditioning on $K=k$ and summing over all possibilities (unconditioning over the distribution of $K$ -- the distribution of  $K$ turns out to be Geometric) will yield
\begin{eqnarray}
\mathbb{E}[Y_{DL}] &=& \sum_{k=1}^\infty \left( \sum_{i=1}^k (\mathbb{E} Y_n^{(i)} + \mathbb{E} Y_{n-1}^{(i)}) \right) p_F (1-p_F)^{k-1} \\
&=&  \sum_{k=1}^\infty \left( \frac{k}{n\lambda} + \frac{k}{(n-1)\lambda + \mu} \right) p_F (1-p_F)^{k-1} \\
&=& \frac{2n\lambda + \mu - \lambda}{n\lambda ((n-1)\lambda + \mu)}  \mathbb{E}[ K] = \frac{2n\lambda + \mu - \lambda}{n\lambda ((n-1)\lambda + \mu)}  \frac{(n-1)\lambda + \mu}{(n-1)\lambda} \\
&=& \frac{\mu + \lambda (2n-1)}{\lambda^2n(n-1)}  \label{c_1MTTDL}
\end{eqnarray}

\begin{figure}[t!]
\centering
\includegraphics[angle=0, height=35mm, width=140mm]{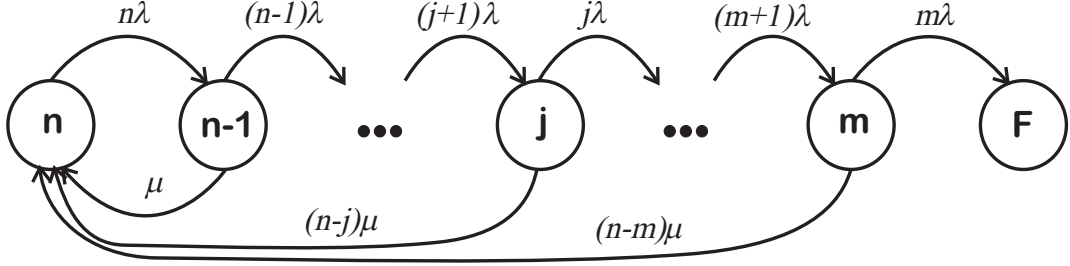}
\caption{A generalized continuous time Markov model for the reliability estimation of $(n,m)$--coded data.}\label{fig:MM4G}
\end{figure}

\hspace{5mm} Note that probability distribution of $Y_{DL}$ is not exponential yet its mean is easy to compute. In fact, the probability distribution of $Y_{DL}$ is a sum of exponentials, but hard to compute it in a closed form. Later, we will see good approximations for $Y_{DL}$ yield a sum of exponentials argument as well.  Let us generalize the Markov model in Fig. \ref{fig:MM4} by letting the system correct $c$ device failures with $m$ data units or devices i.e., $n=m+c$. The previous parity-coded system has $c=1$ and $m=n-1$. Furthermore, let us assume that concurrent repairs are made in each state so that we have the Markov model shown in Fig. \ref{fig:MM4G}. Particularly for this type of generalization, there are Laplace transform-based techniques to compute the MTTDL.

\hspace{5mm} The reliability function $R(t)$ is the probability of being in any state except the state $F$ before time $t$. In otherwords,
\begin{eqnarray}
R(t) = P_n(t) + P_{n-1}(t) + \dots + P_{m+1}(t) + P_{m}(t)
\end{eqnarray}
where $P_j(t)$ is the probability of being in state $j$ at time $t$. Furthermore, Laplace transform of the reliability function $R(t)$ for $t \geq 0$ is given by
\begin{eqnarray}
R^*(s) = \int_{0}^\infty R(t) e^{-st} dt \label{AggregateMTTDL}
\end{eqnarray}

\hspace{5mm}  From equation (\ref{AggregateMTTDL}), we recognize that MTTDL is an aggregate measure of reliability and is given by the evaluation of transformed function at $s=0$, i.e.,
\begin{eqnarray}
MTTDL = R^*(0) = \int_0^{\infty} R(t)dt &=& \int_0^{\infty}  P_n(t)  dt + \dots + \int_0^{\infty} P_{m}(t) dt \\
&=& P_n^*(0) + P_{n-1}^*(0) + \dots + P^*_{m}(0)
\end{eqnarray}

\hspace{5mm} Using some probability theory background, let us write down the conventional Kolomogorov equations in time domain for this model as follows \cite{Rausand},
\begin{eqnarray}
P_n^{\prime}(t) + P_n(t).n\lambda &=& P_{n-1}(t).\mu +  P_{n-1}(t).2\mu + \dots + P_{n-1}(t).(n-m)\mu \nonumber \\
&\dots& \nonumber \\
P_j^{\prime}(t) + P_j(t).(j\lambda + (n-j)\mu) &=& P_{j-1}(t).(j+1)\lambda  \nonumber \\
&\dots& \nonumber \\
P_F^{\prime}(t) &=& P_m(t).m\lambda \nonumber
\end{eqnarray}
and the corresponding transform domain equations shall be,
\begin{eqnarray}
sP_n^*(s) - P_n(0) + P_n^*(s).n\lambda &=& P_{n-1}^*(s).\mu +  P_{n-2}^*(s).2\mu + \dots + P_{m}^*(s).(n-m)\mu \nonumber \\
&\dots& \nonumber \\
sP_j^*(s) - P_j(0) + P_j^*(s).(j\lambda + (n-j)\mu) &=& P_{j-1}^*(s).(j+1)\lambda  \nonumber \\
&\dots& \nonumber \\
sP_F^*(t) - P_F(0) &=& P_m^*(s).m\lambda \nonumber
\end{eqnarray}
with the initial conditions $P_n(0) = 1$ and $P_j(0) = 0$ for $j < n$. These linear set of equations can be put in a matrix form easily as follows,

\[ \left( \begin{array}{cccccc}
s+n\lambda & -\mu & -2\mu & \dots & -(n-m)\mu & 0 \\
-n\lambda & s + (n-1)\lambda + \mu & 0 & \dots & 0 & 0 \\
\vdots & \vdots & \ddots & \vdots & \vdots & \vdots \\
0 & 0 & \dots & 0 & -m\lambda & s \end{array}  \right) \left( \begin{array}{cccccc} P_n^*(s) \\ P_{n-1}^*(s) \\ \vdots \\ P_F^*(s) \end{array}  \right) =
\left( \begin{array}{cccccc} 1 \\ 0 \\ \vdots \\ 0 \end{array}  \right)  \]
and can be expressed in a matrix notation
\[ \mathbf{A} \mathbf{P} = \mathbf{N}^{(0)} \]
from which the inversion of the matrix $(\mathbf{A}^{-1})$ and multiplication from right yields the solution for $\mathbf{P}$ i.e., $P_j^*(s)$ for $j \in \{n, n-1, \dots, m, F\}$ where $\mathbf{N}^{(l)} = [0 \ \ 0 \ \ \dots 0 \ \ 1 \ \ 0 \dots \ \ 0 \ \ 0]$ and $(l+1)$th vector entry is unity. Of course for large $n$, efficient methods to invert the matrix might be needed. Once we find $P_j^*(s)$, using inverse Laplace transform, we can obtain $P_j(t)$ and eventually $R(t)$. For example in \cite{Burkhard}, MTTDL for $c=1$, $c=2$ and $c=3$ are calculated explicitly and they are given by
\begin{eqnarray}
MTTDL_{c=1} &=& \frac{\mu + \lambda(2m+1)}{\lambda^2m(m+1)} \nonumber \\
MTTDL_{c=2} &=& \frac{2\mu^2 + \mu\lambda(5m+6) + \lambda^2(3m^2+6m+2)}{\lambda^3m(m+1)(m+2)} \nonumber \\
MTTDL_{c=3} &=& \frac{6\mu^3 + \mu^2\lambda (17m+33) + \mu\lambda^2 (14m^2+47m+33) + 2\lambda^3(2m^3 + 9m^2 + 11m +3)}{\lambda^4m(m+1)(m+2)(m+3)} \nonumber
\end{eqnarray}

\hspace{5mm} Since we are interested in the first column of the matrix inverse, MATLAB/Mathematica symbolic inverse computation capability may be very useful to calculate explicit expressions for reasonable choices of $c$. As can be seen $MTTDL_{c=1}$ is the same as the equation  in (\ref{c_1MTTDL}) we previously derived. We observe that for large $c$, MTTDL expressions get quite complicated. In conventional storage systems however, $\omega = \mu/\lambda$ ratio is very large. Therefore, simplifications to MTTDL can be performed using the assumption $\omega = \mu/\lambda \gg 1$. In \cite{Walter}, an approximate expression for the generic MTTDL is derived based on the assumption that $\omega$ ratio is very large. The approximate expression is quite neat and given by
\begin{eqnarray}
MTTDL_{c} \approx \frac{\omega^c}{\lambda (n-c) \binom{n}{c}} \label{MTTDLSimple}
\end{eqnarray}

\hspace{5mm} Unfortunately, even with this assumption the expressions for $R(t)$ for large values of $c$ gets very complicated. In \cite{Burkhard}, good approximations for $R(t)$ are given for any $m$ and $c=0,1,2$ and $3$ as follows
\begin{eqnarray}
R_{c=0}(t) &=& e^{-m\lambda t}  \\
R_{c=1}(t) &\approx& C_{1,m,0}e^{-t/MTTDL_{c=1}} + C_{1,m,-\mu}e^{-t(\mu+\lambda(2m+1))} \\
R_{c=2}(t) &\approx& C_{2,m,0}e^{-t/MTTDL_{c=2}} + C_{2,m,-\mu}e^{-t(\mu+\lambda(2m+3))} + C_{2,m,-2\mu}e^{-t(2\mu+\lambda m)} \\
R_{c=3}(t) &\approx& C_{3,m,0}e^{-t/MTTDL_{c=3}} + C_{3,m,-\mu}e^{-t(\mu+\lambda(2m+5))} \nonumber \\
&& \ \ \ \ \ + C_{3,m,-2\mu}e^{-t(2\mu+\lambda(m+1))} + C_{3,m,-3\mu}e^{-t(3\mu+\lambda m)}
\end{eqnarray}
where the coefficients of the exponentials are
\begin{eqnarray}
C_{c,m,0} = 1 + \sum_{i=1}^c \frac{1}{i} m \binom{m+c}{c} \omega^{-c-1}
\end{eqnarray}
and
\begin{eqnarray}
&& C_{1,m,-\mu} = -\frac{\lambda^2m(m+1)}{\mu^2}, C_{2,m,-\mu} = -\frac{\lambda^3m(m+1)(m+2)}{\mu^3}, C_{3,m,-\mu} = -\frac{\lambda^4m(m+1)(m+2)(m+3)}{2\mu^4}, \nonumber \\
&& C_{2,m,-2\mu} = -\frac{\lambda^3m(m+1)(m+2)}{4\mu^3}, C_{3,m,-2\mu} = -\frac{\lambda^4m(m+1)(m+2)(m+3)}{4\mu^4}, \nonumber \\
&& C_{3,m,-3\mu} = -\frac{\lambda^4m(m+1)(m+2)(m+3)}{18\mu^4}. \nonumber
\end{eqnarray}

\hspace{5mm} Through a tedious algebra, these expressions can be generalized to any $m$ and $c$, although the original work has not derived those expressions \cite{Burkhard}. Exact closed form expressions for the general case is an open problem.

\subsection{Durability and MTTDL}

\hspace{5mm}  One of the important metrics related to the theory of reliability is \emph{durability}. This metric is defined to be the duration of time the system is able to provide access to the user data or entity. If the data is protected by a redundancy mechanism i.e., replication or erasure correction coding, durability refers to the permanent loss of the encoded/replicated user data  rather than the permanent loss of parity or replica blocks. The unit of durability is usually expressed in terms of days or years. In industry terminology however, durability is expressed in terms of nines (9's). This refers to the probability of seeing no error during the operation of the system for the first few number of years.

\hspace{5mm} As mentioned before, the aggregate nature of the MTTDL
reveals only very little information about the reliability/durability of the overall system. Exact calculations for the time to data loss  distribution will yield the actual information about the reliability of the system, yet it may be cumbersome to compute it.  We have seen that $R(t)$ is not exponentially distributed in general. However for simplicity let us assume it is distributed exponentially with the rate $1/MTTDL$. The number of 9's is then given by $\left\lfloor \log_{10}\left(\frac{1}{1-R(t)}\right)\right\rfloor$ where $R(t) = e^{-t/MTTDL}$ is the reliability/durability function and $t \geq 0$ is the time expressed in units of MTTDL (usually in hours). For example, for a system with MTTDL of $2,500,000$ hours, and an operating time of interest of 1 year ($8760$ hours), we have the durability number computed as follows,
\begin{eqnarray}
R(8760) = e^{-8760/2500000} = 0.9965 \rightarrow \textrm{two 9's}
\end{eqnarray}
which means the system will operate without a failure with probability $0.9965$ for the first year of use at a $100\%$ duty cycle. Next, this simplistic view will be compared to the  approximations given in the previous section for a replication redundancy scheme. We shall see later in the document that the exponential approximation is pretty good for durability computations.

\subsection{A case study: Replication v.s. Erasure Correction}

\hspace{5mm} Data storage systems use different redundancy
schemes to prevent data loss that can occur because of multiple concurrent\footnote{Here the failures need not to fail at exactly the same point in time. We rather refer to the extra failures as concurrent failures before the repair of the failed devices is completed.}
device failures. Replication is one of the widely used schemes
where each data block is replicated and the replicas are stored
in different nodes to improve the chances that at least one replica survives when multiple storage nodes fail. In a replication scheme, $m=1$ and $c=1,2,\dots$. Table \ref{Table1} illustrates that for large $\omega$, approximations given for $MTTDL$ and the time to failure distribution being exponential are quite accurate, particularly when the quantities are expressed in terms of nines.

\begin{table}[t!]
\begin{center} {\footnotesize
\begin{tabular}{|cc|cc|cc|cc|}
\hline
 \multicolumn{2}{|c|}{Fail/Repair rates} & \multicolumn{2}{c|}{$c=1$} & \multicolumn{2}{c|}{$c=2$} &
\multicolumn{2}{c|}{$c=3$}  \\
$\lambda$ & $\mu$  & \multicolumn{1}{c}{$e^{-t/MTTDL_{1}}$} & \multicolumn{1}{c|}{$R_{c=1}(t)$} &
\multicolumn{1}{c}{$e^{-t/MTTDL_{2}}$} & \multicolumn{1}{c|}{$R_{c=2}(t)$} & \multicolumn{1}{c}{$e^{-t/MTTDL_{3}}$} &
\multicolumn{1}{c|}{$R_{c=3}(t)$} \\\hline
$1/200K$ &     1/24 & 4  &   4 & 8  &  8 & 12 &  12  \\
$1/500K$  &     1/24 & 5  &   5 & 9  &  9 & 14 & 14     \\
$1/1.2M$ &     1/24 & 6  &   6 & 11  &  11 & 15 & 15    \\
$1/200K$    &   1/240 & 3  &   3 & 6  &  6 & 9 & 9     \\
$1/500K$   &    1/240 & 4  &   4 & 7  &  7 & 11  & 11   \\
$1/1.2M$    &   1/240 & 5  &   5 & 9  &  9 & 12  & 12    \\
\hline
\end{tabular} }
\end{center}
\caption{Number of Nines for 3 and 4-replicated systems per archive.}
\label{Table1}
\end{table}

\hspace{5mm} Let us increase the number of data devices from $m=1$ to $m=100$ and recalculate these values for $c=1,2,3$. The results are shown in Table \ref{Table2}. As can be seen, approximations are still mostly accurate, if not, they are an underestimator. 

\begin{table}[t!]
\begin{center} {\footnotesize
\begin{tabular}{|cc|cc|cc|cc|}
\hline
 \multicolumn{2}{|c|}{Fail/Repair rates} & \multicolumn{2}{c|}{$c=1$} & \multicolumn{2}{c|}{$c=2$} &
\multicolumn{2}{c|}{$c=3$}  \\
$\lambda$ & $\mu$  & \multicolumn{1}{c}{$e^{-t/MTTDL_{1}}$} & \multicolumn{1}{c|}{$R_{c=1}(t)$} &
\multicolumn{1}{c}{$e^{-t/MTTDL_{2}}$} & \multicolumn{1}{c|}{$R_{c=2}(t)$} & \multicolumn{1}{c}{$e^{-t/MTTDL_{3}}$} &
\multicolumn{1}{c|}{$R_{c=3}(t)$} \\\hline
$1/200K$ &     1/24 & 1  &   1 & 3  &  3 & 5 &  5  \\
$1/500K$  &     1/24 & 2  &  2 & 4  &  4 & 7 & 7     \\
$1/1.2M$ &     1/24 & 2  &   2 & 5  &  5 & 8 & 8    \\
$1/200K$    &   1/240 & 0  &   0 & 1  &  1 & 2 & 3     \\
$1/500K$   &    1/240 & 1  &   1 & 2  &  2 & 4  & 4   \\
$1/1.2M$    &   1/240 & 1  &   1 & 3  &  3 & 5  & 6    \\
\hline
\end{tabular} }
\end{center}
\caption{Number of Nines for Erasure-coded systems per archive.}
\label{Table2}
\end{table}

\subsection{Discussion on the accuracy of MTTDL}

\hspace{5mm} Although MTTDL is a useful tool for making
relative comparisons, it is meaningless measure for the absolute measurements \cite{Greenan}. In addition, given the unrealistic assumptions used to derive closed-form expressions, it became a subject of question recently. A system designer may be interested in the probability of
failure for the first few years instead of the mean time to failure. The aggregate nature of the MTTDL usually conveys very limited information to the customer. In fact, $R(t)$, also known as durability, is what most of the customers are interested in.

\hspace{5mm} Traditional reliability models were constructed with four simplifying assumptions: the only failures are whole-device failures, the use of single-disk fault-tolerant
codes such as parity coding, devices fail at a constant rate, devices are repaired at a constant rate. While traditional Markov models enable quick analytic reliability
estimates of single-disk fault-tolerant systems, they do not extend well to multi-disk
fault-tolerant systems, the inclusion of sector failures and they do not compensate for
time dependence in failure and repair.

\hspace{5mm}  There are several problems with these set of assumptions of a simple Markov model. For example, many models do not
account for sector failures, do not accurately model device rebuild processes and assume that all
devices exhibit the same failure/rebuild rates. In addition, extension to a general model that accounts for $m$-erasure correction is not straightforward. Canonical Markov models have memoryless interarrival times, yet most of the time, the storage components already in the rebuilt process are assumed to start repair from the beginning if another storage component fails. These considerations led research community to think of more advanced modeling techniques to estimate the reliability more accurately. Most of them emphasize the hardness of deriving analytical results and encourage Monte Carlo type simulations for accurate estimations \cite{GreenanThesis}.

\subsection{Arguments in favor of MTTDL and Advanced models}

\hspace{5mm} Although the inaccuracies of the previous calculations of MTTDL are plenty, this notorious reliability metric, based on exponential failure and repair times, has been shown to be insensitive to the actual distribution of failure/repair times as long as the constituent storage devices have MTTF being much larger than their MTTR \cite{Venkatesan} and operate independent of eachother.  This result essentially implies that the system MTTDL computations of previous sections
will not be affected if the device failure distributions were changed
from an exponential to a some other distribution with the same mean. Given that, there have been few attempts to make the original Markov model more realistic. We shall review them in this section and propose the most general form of the canonical model for reliability estimations.

\begin{figure}[t!]
\centering
\includegraphics[angle=0, height=39mm, width=140mm]{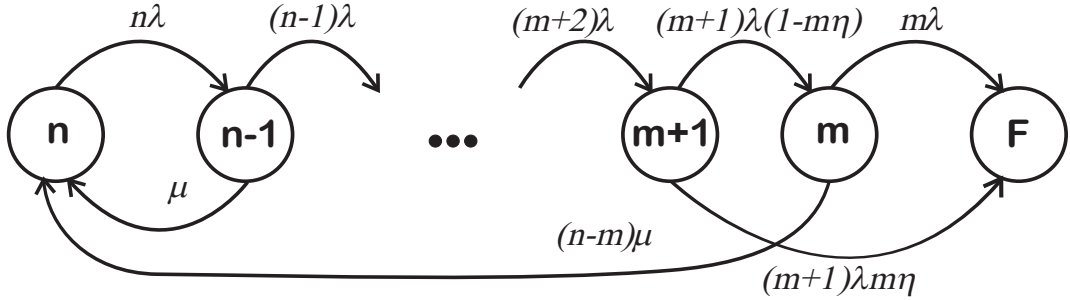}
\caption{A continuous time Markov model for the reliability estimation of $(n,m)$--coded data incorporates the hard errors in $MTTDL$ estimation.}\label{fig:MM4GG}
\end{figure}

\subsubsection{Incorporating a hard error (uncorrectable error)} 

\hspace{5mm}  This requirement is observed to be necessary when the system operates in the critical mode i.e., a state in which one more device failure leads to total system crash and/or data loss. This requirement imposes a slight change in our Markov model as shown in Fig. \ref{fig:MM4GG}. In this model, $\eta$ represents the probability of seeing an uncorrectable error per device read during the rebuilt process.  Let $\verb"UCER"$ denote the uncorrectable error rate of the device (such as $10^{-15}$, expressed in terms of errors per number of bytes or bits read), $\eta$ is given by $\verb"device" \ \ \verb"capacity" \times \verb"UCER"$. Although this is the published expression \cite{Hafner}, it comprises an inaccuracy that $\verb"device" \ \ \verb"capacity" \times \verb"UCER"$ can be greater than 1. However, $\eta$ appears to be a probability term and must satisfy $0 \leq \eta \leq 1$. Thus, a more accurate expression can be given as $\eta = 1 - (1 - \verb"UCER")^{\verb"device" \ \ \verb"capacity" }$.

\hspace{5mm}  The transition from state $m+1$ to state $F$ is introduced to model the rate at which the system encounters an uncorrectable error while reading and rebuilding $t$ device failures. Based on the analysis given in \cite{Hafner}, the rate is computed as the product of the rate that a disk fails when $(m+1)$ devices are available, i.e., $(m+1)\lambda$ and the probability of encountering an uncorrectable error when reading $m$ devices for rebuild (Note here that we assume conventional MDS codes, which may require many device reads), i.e., $P_{UCER} = m\eta$. Similar to our previous argument, this product can assume values greater than one. A more accurate expression shall be given by
\begin{eqnarray}
P_{UCER} = 1 - (1 - \eta)^m \label{PUCER}
\end{eqnarray}

\hspace{5mm} An approximate solution for the MTTDL (without the corrections given for $\eta$ and $P_{UCER}$ i.e., $\eta = \verb"device" \ \ \verb"capacity" \times \verb"UCER"$ and $P_{UCER} = m\eta$) is given in \cite{Hafner} for one-device-at-a-time repair strategy. Using similar arguments we can easily extend those results to concurrent repairs and obtain the following expression for $\eta \ll n$ and $\omega = \mu/\lambda \gg 1$,
\begin{eqnarray}
MTTDL_c \approx \frac{c! w^c}{n(n-1)\dots(n-c)(\lambda+\eta\mu)}
\end{eqnarray}

\begin{figure}[t!]
\centering
\includegraphics[angle=0, height=39mm, width=140mm]{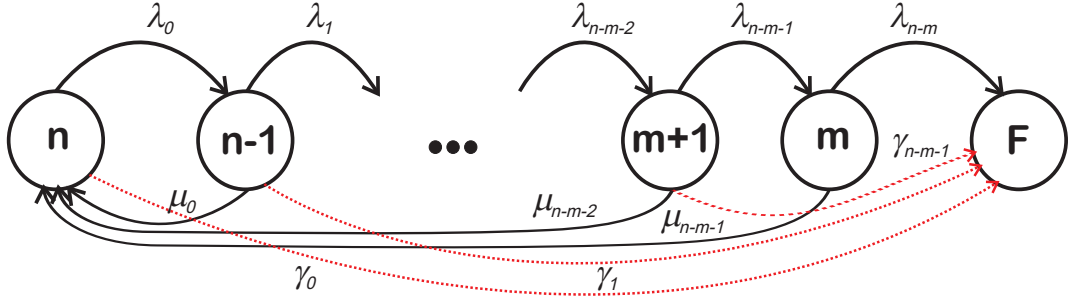}
\caption{A generalized continuous time Markov model that can be used for the reliability estimation of $(n,m)$--coded data that incorporates the irregular fault tolerance as well as the hard errors in $MTTDL$ estimation.}\label{fig:MM4GX}
\end{figure}

\subsubsection{Modifications for Irregular Codes Or multiple regular codes} 

\hspace{5mm} In addition to previous improvements, let us generalize the model to cover the class of more popular \emph{irregular} erasure correction codes. By ``irregular", we mean the modern non-MDS erasure correction codes that are constructed based on bipartite graphs and has irregular fault tolerance. Most prominent ones of this class are XOR-based LDPC or fountain codes that possess good reparability properties with simple encoding and decoding operations.

\hspace{5mm} Let us introduce one of the most general Markov model and its application to reliability analysis of systems which are protected by irregular codes. Note that few generalizations are made in \cite{Hafner}, but not to that level shown in Fig. \ref{fig:MM4GX}. Furthermore, some special cases of this model is considered and analyzed such as in \cite{Paris}. In this model, device failures happen one at a time with rate $\lambda_i$ while due to irregular fault tolerance of the erasure correction code, it is possible to go from any state except state $m$ to fail state with some rate $\gamma_i$. Also, as mentioned in the past research that Markov models assume rebuild clock is ignored. In other words,  each Markov model resets the rebuild
time for all disks being rebuilt whenever another disk fails during rebuild. The model of
concurrent rebuilds considered in this document adopts a rebuild policy that restarts the rebuild of all failed disks
each time a disk fails. Alternatively, one could estimate the remaining repair time at each
state by steadily increasing the rate of ongoing rebuild operations as more failures occur. Thus, in a general model we should allow different rebuild rates $\mu_i$ to meet this goal of better modeling the real world.

\hspace{5mm} With regard to this general model, there are three questions to be explored,
\begin{itemize}
\item Given the nature of the irregular code and its fault tolerance, how do we efficiently compute rates $\{\lambda_i,\gamma_i\}$ to be able to use with this model?
\item Given the nature of rebuilds and associated system format/structure, how do we estimate rates $\{\mu_i\}$ to be able to use with this model?
\item Is there a closed form expression for MTTDL? and $R(t)$? for any $c$ and $m$.
\end{itemize}

\hspace{5mm} \textbf{Let us start with the first question.} This question is in fact answered in \cite{Hafner}. Given the irregular code and the symbol allocation policy (such as CRUSH \cite{CRUSH}), let us denote the probability $p_k$ that the overall system can tolerate one more device failure provided that it has already tolerated $k$ failed devices. There are two scenarios that can lead from an operational (survival) state to failure (absorbing) state. First of all, at the $j$-th state ($n \geq j \geq m$) one more device fails and the correction code cannot tolerate for this loss and the state change ends up with the failure state. The rate of this happening is $j\lambda(1-p_{n-j})$, direct multiplication follows from basic Markov models in probability theory. Another scenario that ends up in data loss is that an extra device fails, the erasure code is able to tolerate. Yet, during the rebuild process a hard error (uncorrectable error) is encountered and the erasure code cannot tolerate this additional error. The rate of this happening is given by $j\lambda p_{n-j}(1-p_{n-j+1})(j-1)\eta$ which can be obtained using conditional probability arguments. Ofcourse, here we assume all the operational $j-1$ devices are used/accessed in the rebuilt process which may not necessarily be the case for regenerating erasure codes \cite{Dimakis}. Thus, the expression can be modified based on the properties of the erasure code used. Finally since $\lambda_{n-j} + \gamma_{n-j} = j\lambda$, for $n \geq j \geq m+1$ we have the following
\begin{eqnarray}
\gamma_{n-j} &=& j\lambda \left( (1-p_{n-j}) + p_{n-j}(1-p_{n-j+1})(j-1)\eta \right) \\
\lambda_{n-j} &=& j\lambda - \gamma_{n-j}
\end{eqnarray}
with $\lambda_{n-m} = (n-m) \lambda$ and $\gamma_{n-m} = 0$.

\hspace{5mm} Here, we still need to explain how to obtain conditional probabilities $p_k$s. Let us denote the unconditional probability $q_k$ that the overall system can tolerate $k$ device failures out of $\binom{n}{k}$ total number of possible instances. Suppose that $s_k$ of these possibilities can be tolerated by the system, then we have $q_k = s_k\bigg/\binom{n}{k}$. Suppose that a $k+1$ pattern of failures the system can tolerate, than any failure subset of size $k$ of these $k+1$ failures can be tolerated. This means that $p_k = q_{k+1}/q_k$, or more explicitly,
\begin{eqnarray}
p_k = \frac{s_{k+1}}{\binom{n}{k+1}} \Big/ \frac{s_{k}}{\binom{n}{k}} = \frac{s_{k+1}(k+1)}{s_{k}(n-k)}
\end{eqnarray}
where $s_k$s are usually found through an elaborate test of the overall system (erasure code performance/properties plus the symbol allocation strategies) protected by the irregular erasure code. Independent of the allocation policy, studies like \cite{HPpaper} investigate fault tolerance metrics such as minimum erasure patterns for any linear block code. The results of such studies can be useful for the computation of durability numbers of erasure-coded storage systems, as we shall see shortly. 

\hspace{5mm} \textbf{For the second question,} we can simply assume $\mu_i = \mu$ (namely a homogenous repair process) where the rebuild time for $n-i$ devices is the same as the rebuild times for a single derive. Ofcourse more elaborate approaches will increase the accuracy in the final durability results. In another system,  concurrent failure repair may imply $\mu_i = (i+1)\mu$ (namely progressive repair) because each device is repaired with the same rate $\mu$. Thus, the choice for repair rates are a strong function of the overall system implementation.

\hspace{5mm} \textbf{As for the third question}, we can use Mathematica or Matlab symbolic toolbox to calculate closed form expressions for small values of $m$ and $c$. The derivation of a closed form expression for any $m$ and $c$ can be significant for two main reasons. First, we can model the durability modeling for long block length irregular codes (whose block sizes are practically designed to be long compared to that of conventional algebraic codes). Secondly, we can model multiple (in fact a vast array of) short length MDS-protected RAID type systems.

\begin{figure}[t!]
\centering
\includegraphics[angle=0, height=39mm, width=140mm]{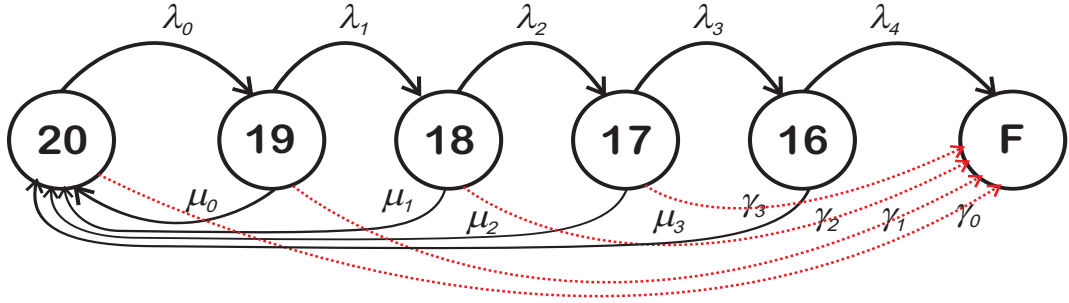}
\caption{A generalized Markov model used for the reliability estimation of two $(10,8)$ RAID 6 arrays.}\label{fig:MM4twoRAID6}
\end{figure}

\hspace{5mm} Let us give an example for the latter: consider  two 2 out-of 10 ($m=8$, $c=2$) RAID6 systems that are used to store user data with $n=20$. We notice that
\begin{itemize}
\item Every zero-device failure  can be tolerated ($q_0 = 1$).
\item Every one-device failure (each ${ 20 \choose 1 } = 20$ possibilities) can be tolerated ($q_1 = 1$).
\item Every two-device failures (each ${ 20 \choose 2 } = 190$ possibilities) can be tolerated ($q_2 = 1$).
\item Only 900 three-device failures out-of $1140 = { 20 \choose 3 }$ can be tolerated ($q_3 = 0.7895$).
\item Only 2025 three-device failures out-of $4845 = { 20 \choose 4 }$ can be tolerated ($q_4 = 0.4180$).
\item No device failure larger than four can be tolerated ($q_k = 0$ for $k>4$).
\end{itemize}
from which we deduce $\{p_0,p_1,p_2,p_3, p_4\} = \{1, 1, 0.7895, 0.5294, 0\}$. The Markov model shown in Fig. \ref{fig:MM4twoRAID6} is used to estimate the reliability with the following set of parameters,
\begin{itemize}
\item $\gamma_0$ = $n\lambda \left((1-p_0) + p_0(1-p_1)(n-1)\eta\right) = 0$, \\ $\lambda_0 = n\lambda - \gamma_0 = n\lambda = 20\lambda$.
\item $\gamma_1$ = $(n-1)\lambda \left((1-p_1) + p_1(1-p_2)(n-2)\eta\right) = 0.2105 \times (n-1)(n-2)\lambda \eta = 72 \lambda \eta$, \\ $\lambda_1 = (n-1)\lambda - \gamma_1 = 19 \lambda - 72 \lambda \eta$.
\item $\gamma_2$ = $(n-2)\lambda \left((1-p_2) + p_2(1-p_3)(n-3)\eta\right) = 3.79\lambda - 33.4\lambda \eta$, \\ $\lambda_2 = (n-2)\lambda - \gamma_2 = 18\lambda - 3.79\lambda  + 33.4\lambda \eta = 14.21\lambda + 33.4\lambda \eta$.
\item $\gamma_3$ = $(n-3)\lambda \left((1-p_3) + p_3(1-p_4)(n-4)\eta\right) = 8\lambda - 144\lambda \eta$ \\
 $\lambda_3 = (n-3)\lambda - \gamma_3 = 17 \lambda - 8\lambda + 144\lambda \eta = 9\lambda + 144\lambda \eta$.
\item $\gamma_4$ = 0 \\ $\lambda_4 = (n-4)\lambda =16 \lambda $.
\end{itemize}

\begin{table}[t!]
\begin{center} {\footnotesize
\begin{tabular}{|cc|cc|cc|cc|}
\hline
 \multicolumn{2}{|c|}{Fail/Repair rates} & \multicolumn{2}{c|}{General Markov (HR)}  & \multicolumn{2}{c|}{General Markov (PR)} & \multicolumn{2}{c|}{Conventional Method  (PR)}   \\
$\lambda$ & $\mu$  & MTTDL & Durability 9's  & MTTDL & Durability 9's  & MTTDL & Durability 9's  \\\hline
$1/200K$ &     1/24 & $1.035 \times 10^9$  &   5 & $1.1 \times 10^9$ & 5  & $1.93 \times 10^{10}$ & 6 \\
$1/500K$  &     1/24 & $6.9 \times 10^9$   &  5  & $7.1 \times 10^9$ & 5   & $3.01 \times 10^{11}$ & 7   \\
$1/1.2M$ &     1/24 & $4.1 \times 10^{10}$   &   6  & $4.13 \times 10^{10}$ & 6  & $4.17 \times 10^{12}$ & 8  \\
\hline
\end{tabular} }
\end{center}
\caption{Number of nines for two RAID6 systems with $m=8$ and $c=2$. HR: Homogenous Repair i.e., $\mu_i = \mu$, PR: Progressive Repair i.e., $\mu_i = (i+1)\mu$.}
\label{Table3}
\end{table}

\hspace{5mm} Let us further assume we use disks of size $1TB$ as our storage devices with $10^{-15}$ uncorrectable error rate i.e., $\eta = 10^{-3}$. The results for MTTDL as well as durability in terms of $9's$ are obtained using symbolic computations through MATLAB and are evaluated/presented in Table \ref{Table3}. Closed form expressions for this general case, $m$ and $c$ to allow efficient computation will be given in the next section.

\hspace{5mm} As can be seen changing the repair strategy has only minor effect on the MTTDL results. Also included are the set of results using the basic model for each RAID6 array and compute the MTTDL using Equation (\ref{MTTDLSimple}). It assumes exponential data loss probability distribution and thus computes the MTTDL of the two RAID6 arrays based on a simple algebraic addition of data loss rates. This conventional scheme does not take into account the uncorrectable errors as well. As expected, the results with conventional scheme show more optimistic MTTDL and durability numbers.

\hspace{5mm} \emph{For an arbitrary number of MDS-protected arrays of devices}, it turns out that there is a closed form expression for $s_k$. In other words, let us assume we have $\pi$ arrays each having $n$ devices $c$ of which are redundant for failure recovery. The total number of choosing $k$ failed devices out of $\pi n$ is given by $\binom{\pi n}{k}$. Let us assume further that there are $r_{j}$ number of arrays with $j$ device failures with $0 \leq j \leq c$.  The general expression for $s_k$ can be given by \cite{Hafner}
\begin{eqnarray}
s_k = \sum_{r_0 + r_1 + \dots + r_{c} = \pi \atop \sum_{i=0}^c ir_i = k} \binom{\pi}{r_0, r_1, \dots, r_{c}} \prod_{i=0}^c \binom{n}{i}^{r_i}
\end{eqnarray}
This expression follows because
\begin{itemize}
\item the multinomial coefficient counts the number of combinations we can distribute $i$ failures over $r_i$ arrays etc.
\item Given that $r_i$ arrays with $i$ failures, there are $\binom{n}{i}^{r_i}$ ways for distributing those failures,
\end{itemize}
with the constraint that total number of arrays must be $\pi$ and failures be $k$. Multinomial theorem is a generalization of the binomial theorem. This generalized version implies that we have
\begin{eqnarray}
(x_0 + x_1 + \dots + x_c)^\pi = \sum_{r_0 + r_1 + \dots + r_c = \pi} \binom{\pi}{r_0, r_1, \dots, r_{c}} \prod_{i=0}^c x_i^{r_i}
 \end{eqnarray}

 \hspace{5mm} If we replace $x_i$ with $\binom{n}{i}x^i$, we shall have
 \begin{eqnarray}
\left(1 + \binom{n}{1}x + \binom{n}{2}x^2 + \dots + \binom{n}{c}x^c\right)^\pi = \sum_{r_0 + r_1 + \dots + r_c = \pi} \binom{\pi}{r_0, r_1, \dots, r_{c}} \prod_{i=0}^c \binom{n}{i}^{r_i} x^{\sum_{i=0}^c ir_i}
 \end{eqnarray}
from which we notice that if we set $\sum_{i=0}^c ir_i =k$, we can realize that $s_k$ is the coefficient in this expansion for $x^{k}$, i.e.,
\begin{eqnarray}
s_k = coef\left(\left(\sum_{i=0}^c\binom{n}{i}x^i\right)^\pi, x^k\right) \label{skforarray}
\end{eqnarray}

\hspace{5mm} Therefore, we have
\begin{eqnarray}
p_k = \frac{coef\left(\left(\sum_{i=0}^c\binom{n}{i}x^i\right)^\pi, x^{k+1}\right) (k+1)}{coef\left(\left(\sum_{i=0}^c\binom{n}{i}x^i\right)^\pi, x^{k}\right) (\pi n - k) }
\end{eqnarray}

\begin{figure}[t!]
\centering
\includegraphics[angle=0, height=60mm, width=125mm]{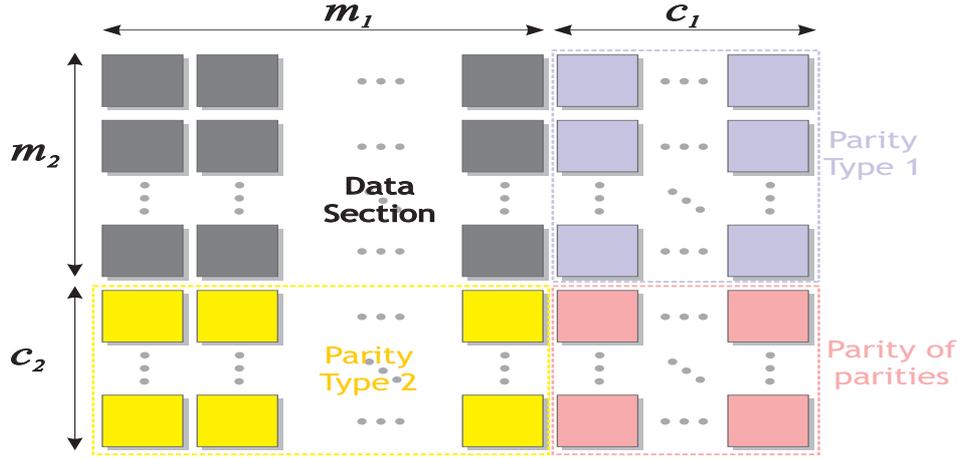}
\caption{A two-dimensional RAID scheme where $m_1 m_2$ data blocks are protected by a total of $c_1 k_2 + c_2k_1 + c_1c_2$ parity blocks.}\label{fig:MM4twoRAID6}
\end{figure}

\subsubsection{2D storage arrays using MDS codes} 

\hspace{5mm} Another interesting case is the use of MDS codes in the context of two dimensional storage arrays. In the past, particular class of two-dimensional RAID schemes are shown to provide better reliability against single dimensional counterparts \cite{Paris2}. However, no general treatment has been conducted. There are recent studies like \cite{Suayb2}, they only consider the mean time to the first failure of multidimensional RAID type storage systems using certain assumptions like independence. Here our consideration is more general for two-dimensional arrays and aims at finding reliability estimates about MTTDL using the general Markov model. For this, we need to determine $s_k$s and error tolerability of two dimensional storage arrays. We assume that each horizontal  and  vertical array component are protected by an MDS code with $c_1$ and $c_2$ parity blocks, respectively. A simple bounded distance decoder is assumed. Decoding can start either in horizontal or vertical direction wherever appropriate. For notation simplicity let $n_1 = m_1+c_1$ and $n_2 = m_2 + c_2$.  We have the following conjecture for $s_k$ if $k \leq (c_1+1)(c_2+1) + \min\{c_1,c_2\}$

\[
 s_k =
  \begin{cases}
   \binom{n_1n_2}{k} & \text{if } k \leq c_1c_2 + c_1 + c_2 \\
   \binom{n_1n_2}{k} - \binom{n_1}{c_1+1}\binom{n_2}{c_2+1}\binom{n_1n_2 - (c_1+1)(c_2+1)}{ k - (c_1+1)(c_2+1) }      & \text{if } (c_1+1)(c_2+1) \leq k \leq (c_1+1)(c_2+1) + \min\{c_1,c_2\}
  \end{cases}
\]
\hspace{5mm} The exact expressions for $s_k$ for $k > (c_1+1)(c_2+1) + \min\{c_1,c_2\}$ gets more complicated and the closed form expressions for any $k$ is an open problem. Let us consider an example that is a special case of the 2D storage arrays using MDS codes.

\hspace{5mm} \textbf{Example 1:} \emph{A mirrored $(n_1,m_1)$-coded array ($c_1 = n_1 - m_1$ parities)}, is a special case of a 2D storage array where the vertical encoding is nothing but a single parity generation though information duplication. For example RAID 51 and RAID 61 systems are special cases of this example. In order for a column to fail, both of the copies must be unavailable/failed. For a given $k$ device failures in the 2D array, there are at most $\lfloor k/2 \rfloor$ column failures. Since the horizontal array can tolerate $c_1$ device failures at most, maximum number of column failures can not exceed $\min\{\lfloor k/2 \rfloor, c_1\}$ in order for the overall system be able to recover the data. For a given $0 \leq j \leq \min\{\lfloor k/2 \rfloor, c_1\}$ column failures, we realize that equation $(\ref{skforarray})$ can be used with $k = k - 2j$, $n=2$ and $r = n_1$ to calculate number of tolerable cases for each $j$. Note also that each case constitute a disjoint set. Since we have $\binom{n_1}{j}$ possible selections of these columns, we have
\begin{eqnarray}
s_k = \sum_{j=0}^{\min\{\lfloor k/2 \rfloor, c_1\}} \binom{n_1}{j} coef\left(\left(1 + 2x\right)^{n_1}, x^{k-2j}\right) \textbf{1}_{y \leq n_1 +j}(k)
\end{eqnarray}
where $\textbf{1}_{y \leq A}(y)$ is the indicator function that evaluates to 1 if $y \leq A$. This function is needed in our expression because for each term in the summation, we assume there are $j$ column failures and  if $k - 2j > n_1 - j$ or $k > n_1 + j$, it would mean that there are at least $j+1$ column failures, which is contrary to our conditional statement that we have $j$ column failures. 

\begin{figure}[t!]
\centering
\includegraphics[angle=0, height=65mm, width=100mm]{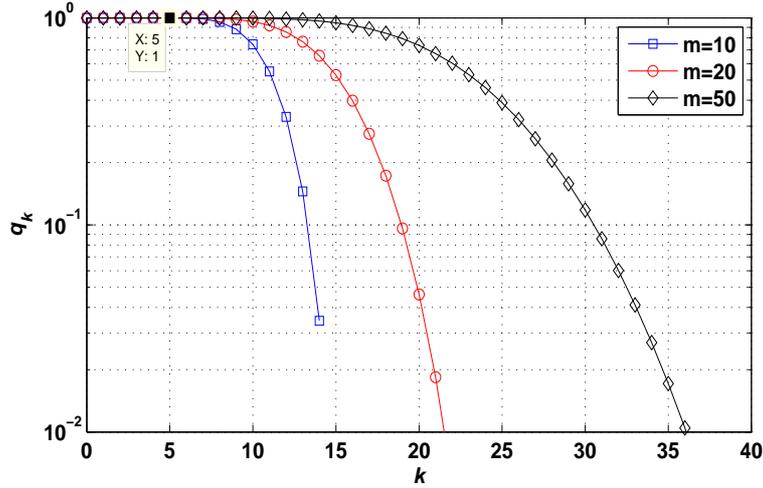}
\caption{ $q_k$ values as a function of $k$ and $m$ for a RAID61 configuration.}\label{fig:RAID61}
\end{figure}

\hspace{5mm} In Fig. \ref{fig:RAID61}, the computation for $q_k$ is shown for RAID61 ($c_1 = 2$) configuration with $m_1 = 10,20,50$ and $n_2=2$ as a function of $k$. Notice that for each case if $k \leq c_1c_2 + c_1 + c_2 = 5$, all combinations of $k$ failed devices are tolerable. It might be interesting for example to drive a similar expression for $n_2=3$ while we keep $c_2=1$ by using parity coding. Even in this simple extension, the expression can get pretty complex. We also note that the precise definition of the decoding algorithm across all devices also play a key role for the calculation of $s_k$ for this particular configuration. For instance, vertical and horizontal decoding iterations can be allowed until the decoding can no longer improve the failure statistics.

\hspace{5mm} \textbf{Exercise 1:} Using the result of the above example, compute the MTTDL estimates of RAID51, RAID6 and RAID61 for comparison. Use $\lambda = 1/200K$ and $\mu = 1/24$ in your computations.

\hspace{5mm} Using this result, a generalization to any $n_2$ with $m_2=1$ and $c_2 = n_2 - 1$ can easily be made. The result of this generalization gives a closed form expression as indicated in the following lemma.

\hspace{5mm} \textbf{Lemma 2:} \emph{For a 2D array of storage devices with only one constraint that $m_2 = 1$, the closed form expression for $s_k$ is of the form
\begin{eqnarray}
s_k = \sum_{j=0}^{\min\{\lfloor k/n_2 \rfloor, c_1\}} \binom{n_1}{j} coef\left(\left((1+x)^{n_2} - x^{n_2}\right)^{n_1}, x^{k-n_2j}\right) \textbf{1}_{y \leq n_1 +c_2 j}(k)
\end{eqnarray}
%\begin{eqnarray}
%\xi_t = \gamma_{t-3} \mu_{t-3} \prod_{i=0}^{t-4}\lambda_i
%\end{eqnarray}
}

\hspace{5mm} PROOF: Proof of this lemma is pretty straightforward using the previous arguments of the example where $c_2=1$ and $n_2 =2$. \hfill $\square$

\hspace{5mm} The closed form expressions for $s_k$ without the constraint of lemma 2 becomes more complicated. This is in fact the same interesting combinatorial problem, mentioned earlier. Although many studies have been conducted regarding ball and bins argument in the past to be used to calculate such quantities, with the specific decoding  principle we assume in this framework (iterative both in horizontal and vertical directions), there appears to be no solution for the general case published yet.

%%%%%%%%%%%%%%%%%%%%%%%%%%%%%%%%%%%%%%%%%%%%%%%%%%
%%%% LEFT HERE
%%%%%%%%%%%%%%%%%%%%%%%%%%%%%%%%%%%%%%%%%%%%%%%%%%
%%%%%%%%%%%%%%%%%%%%%%%%%%%%%%%%%%%%%%%%%%%%%%%%%%
%%%%%%%%%%%%%%%%%%%%%%%%%%%%%%%%%%%%%%%%%%%%%%%%%%
%%%%%%%%%%%%%%%%%%%%%%%%%%%%%%%%%%%%%%%%%%%%%%%%%%
%%%%%%%%%%%%%%%%%%%%%%%%%%%%%%%%%%%%%%%%%%%%%%%%%%
%%%%%%%%%%%%%%%%%%%%%%%%%%%%%%%%%%%%%%%%%%%%%%%%%%
%%%%%%%%%%%%%%%%%%%%%%%%%%%%%%%%%%%%%%%%%%%%%%%%%%
%%%%%%%%%%%%%%%%%%%%%%%%%%%%%%%%%%%%%%%%%%%%%%%%%%
%%%%%%%%%%%%%%%%%%%%%%%%%%%%%%%%%%%%%%%%%%%%%%%%%%

\subsubsection{XOR-based erasure codes} 

\hspace{5mm} \emph{For an arbitrary erasure code -- possibly non-MDS protected array of devices}, the computation of $s_k$ is challenging and is a strong function of the code's graphical construction. Even without the consideration of coded symbol allocation methodology, identification of code fault tolerance might be quite complex to compute. For example a methodology is presented for a linear code (an XOR-based code) constructed using a systematic binary generator matrix in \cite{Wylie} used for erasure correction. The same algorithm turns out to be applicable to non-systematic codes as well. The claims of that paper are based on the decodability of the code, i.e., the study assumes an optimal decoder. In fact, such a fault tolerance profile is a function of the encoding and decoding algorithms that the code is used with.

\hspace{5mm} For a complete characterization of the fault tolerance of an XOR-based erasure code, every set of erasure patterns must be enumerated. Unfortunately, there are exponentially many such erasure patterns to enumerate. Instead of finding all erasure patterns that lead to decoder failure, \emph{minimal erasures} are found to characterize the fault tolerance of the erasure code. A minimal erasure is a set of erasures that leads to irrecoverable data loss and in which every erasure is both necessary and sufficient for this to be so. Furthermore, an enumeration of all minimal erasures are termed as \emph{minimal erasures list} (MEL). The MEL characterizes completely the fault tolerance of an erasure code while having relatively small size relative to all erasure patterns causing a data loss. An algorithm\footnote{A Matlab implementation of this algorithm can be found at http://www.suaybarslan.com/mea.txt} is presented in \cite{Wylie} for efficiently determining the MEL of a linear XOR-based erasure code defined by a binary generator matrix. Given the MEL of an $(n,k)$ linear XOR-based code, the minimum erasure vector (MEV) is of size $n-k$ in which $i$th entry counts the number of minimum erasure patterns of weight $i$ for $1 \leq i \leq n-k$. It is easy to realize that MEV can be instrumental for the computation of $s_k$.

\hspace{5mm} \textbf{Example 2:} Let us consider the following binary generator matrix for a $(8,4)$ systematic XOR-based linear code,
\[ \textbf{G} =  \left( \begin{array}{cccccccc}
1 & 0 & 0  & 0  & 1 & 0 & 0 & 1  \\
0 & 1 & 0  & 0  & 1 & 1 & 1 & 1  \\
0 & 0 & 1  & 0  & 0 & 1 & 1 & 0  \\
0 & 0 & 0  & 1  & 0 & 0 & 1 & 1 \end{array} \right) \]

\hspace{5mm} Using the ME algorithm, we can compute the MEV = $\{0,0,4,5\}$. From this, we can see that $s_3 = \binom{8}{3} -  4 = 52$ and through simple enumeration $s_4 = \binom{8}{4} - 25 = 45$ and $s_k = 0$ for all $k \geq 5$. 

%\hspace{5mm} If the erasure code performance should be considered in conjunction with the allocation policy, an experimentation can be used to compute values for $s_k$ in a %controlled test environment. However, the exhaustive search of erasure patterns that cause data loss and the frequency of these error patterns is quite complex. Therefore, %efficient methods are needed and shall be the subject of open problems of future investigation.

\section{MTTDL for the Generalized Markov Model}

\hspace{5mm} In this section, we will give the exact as well as approximate expressions for the MTTDL of the most general Markov model introduced in the previous section. The content that we present in this section is partially published in \cite{arslanrel2019}. As we mentioned before, for large  values of $c$, $m$ and large volumes of storage arrays make the conventional computations useless. Closed form expressions are golden in that sense to predict future reliability estimates for long and large size of storage arrays as well as gigantic arrays of data coded by irregular modern coding techniques.

\subsection{Efficient Computation of MTTDL}

\hspace{5mm} Following our previous construction, we have the transform domain Kolomogorov equations put into a coefficient matrix $\textbf{A}$ as shown below.
\small
 \[\mathbf{A} = \left[ \begin{array}{cccccc}
s + \lambda_0 + \gamma_0   & -\mu_0 & -\mu_1 & \dots & -\mu_{c-1} & 0 \\
-\lambda_0 & s + \lambda_1 + \mu_0 + \gamma_1  & 0 & \dots & 0 & 0 \\
0 & -\lambda_1 & s + \lambda_2 + \mu_1 + \gamma_2 & \dots & 0 & 0 \\
0 & 0  & - \lambda_2 & \dots & 0 & 0 \\
\vdots & \vdots &  & \ddots & 0 & 0 \\
0 & 0 & \dots &  s + \lambda_{c-1} + \mu_{c-2} + \gamma_{c-1} & 0  & 0 \\
0 & 0 & \dots &  -\lambda_{c-1} & s + \lambda_{c} + \mu_{c-1}   & 0 \\
-\gamma_{0} & -\gamma_{1} & \dots & -\gamma_{c-1} & -\lambda_{c} & s \end{array} \right]\]
\normalsize

\hspace{5mm} As can be seen, the inversion of $\textbf{A}$ might be quite challenging using symbolic software toolboxes if the size of the matrix is large. Instead, neat closed form expressions are prefered using the special structure of the coefficient matrix. Let us start by stating the key result of this section without proof. For a given number of redundant $c$ units/devices and $x=0,1,\dots,c$, let us define  the following \emph{key rate vector} with the corresponding entries,
\[ \boldsymbol{\Lambda}_x^{(c)} \triangleq \left[ \begin{array}{c}
\lambda_0  \\
\lambda_1  \\
\lambda_2   \\
\vdots   \\
\lambda_{c-1} \\
\lambda_{c}
\end{array} \right] +
\mathbf{V}_x^{(c)}  \left[ \begin{array}{c}
\gamma_0 \\
\mu_0  + \gamma_1   \\
\mu_1  + \gamma_2  \\
\vdots   \\
\mu_{c-2} + \gamma_{c-1} \\
\mu_{c-1}
\end{array} \right]\]
where
\[ \ \ \ \ \ \ \ \ \ \ \ \ \ \   \mathbf{V}_x^{(c)}   = \left[ \begin{array}{cc}
\mathbf{0}_{x} & \mathbf{0}_{x \times c+1-x}   \\
\mathbf{0}_{c+1-x \times x} & \mathbf{I}_{c+1-x}
\end{array} \right]\]
and  $\mathbf{I}_{c+1-x}$ and $\mathbf{0}_{x}$ are identity  and all-zero matrices, respectively.
 Let us assume $\Lambda_x^{c}(j)$ to denote the $(j+1)$-th entry of the vector $\boldsymbol{\Lambda}_x^{(c)} $ for $0 \leq j \leq c$. Then, we have the following transform domain expressions evaluated at $s=0$,
\begin{eqnarray}
P_{m+c-x}^*(0) = \frac{1}{\phi_c(0)}\prod_{\substack{j=0 \atop j \not=x}}^{c}\Lambda_x^{c}(j)
\end{eqnarray}
where the denominator can be computed recursively by
\begin{eqnarray}
\phi_t(0) = \prod_{i=0}^{t}\lambda_i + (\mu_{t-1} + \lambda_t)  \left[ \phi_{t-1}(0) - \prod_{i=0}^{t-1}\lambda_i  + \gamma_{t-1} \left( \prod_{i=0}^{t-2} (\gamma_i + \lambda_i) + \xi_t \right)  \right]\label{Phip}
\end{eqnarray}
for $1 \leq t \leq c$ with the initial condition $\phi_0(0)= \lambda_0$. Here, we can show that $\xi_1 = \xi_2 = 0$ and $\xi_3 = \gamma_0 \mu_0$.  The exact expressions for $\xi_{\{t>3\}}$ get more complicated for large $t$. However, we realize that $\xi_s$s are relatively small compared to the rest of the expression given for $\phi_t(0)$. Thus, a good approximation is
\begin{eqnarray}
\phi_t(0) \approx \prod_{i=0}^{t}\lambda_i + (\mu_{t-1} + \lambda_t)  \left[ \phi_{t-1}(0) - \prod_{i=0}^{t-1}\lambda_i  + \gamma_{t-1} \left( \prod_{i=0}^{t-2} (\gamma_i + \lambda_i) \right)  \right]\label{Phip2}
\end{eqnarray}

Using the above arguments, we can find a closed form expression for $MTTDL_c$ expressed as follows,
\begin{eqnarray}
MTTDL_c = \sum_{x=0}^{c}  P_{m+c-x}^*(0) =  \frac{1}{\phi_c(0)} \sum_{x=0}^{c}  \prod_{\substack{j=0 \atop j \not=x}}^{c}\Lambda_x^{c}(j) \label{MTTDLexact}
\end{eqnarray}

\hspace{5mm} The equation $(\ref{MTTDLexact})$ will become approximate (in fact an overestimator) if we use equation $(\ref{Phip2})$ to approximate $\phi_c(0)$. For example,  we executed the actual computation and the approximation for $\lambda =1/200K$, $\mu=1/24$, $m=8$, $\eta=10^{-3}$, $c=\{1,2,3,4,5\}$ and $r=\{1,2,3\}$ the error due to approximation is observed to be less than $10^{-6}$. We project to find tight bounds on the evolution of $\xi_t$ and predict the maximum error that can be inserted into our computations. For a special case, $\xi_t$ can be expressed in a closed form as stated in the following lemma.

\hspace{5mm} \textbf{Lemma 3:} \emph{For $t \geq 4$  and $x \leq c-2$, if multiple of MDS protected arrays or a XOR-based erasure coded storage array can tolerate any $x$ or less device failures, we have $\gamma_0 = 0 , \dots, \gamma_{x-2} = 0$ and therefore
\begin{eqnarray}
\xi_t \ \ \begin{cases} = \gamma_{t-3} \mu_{t-3} \prod_{i=0}^{t-4}\lambda_i  &\mbox{if } t \leq x+2 \\
 >  \gamma_{t-3} \mu_{t-3} \prod_{i=0}^{t-4}\lambda_i & \mbox{Otherwise.}  \end{cases}
\end{eqnarray}
%\begin{eqnarray}
%\xi_t = \gamma_{t-3} \mu_{t-3} \prod_{i=0}^{t-4}\lambda_i
%\end{eqnarray}
}

\hspace{5mm} PROOF: Proof of this lemma will be provided with when we establish the the evolution of $\xi_t$ as a function of failure and repair rates of the underlying Markovian process. \hfill $\square$

%\hspace{5mm} \textbf{Lemma 2:} \emph{For $c \geq t \geq 4$, if a multiple of MDS protected arrays or a graph based coded storage array can tolerate any $c-2$ or less device failures, we have $\gamma_0 = 0 , \dots, \gamma_{c-4} = 0$ and therefore
%\begin{eqnarray}
%\xi_t = \gamma_{t-3} \mu_{t-3} \prod_{i=0}^{t-4}\lambda_i
%\end{eqnarray}}

%\hspace{5mm} PROOF: Proof of this lemma will be provided with when we establish the the evolution of $\xi_t$ as a function of failure and repair rates of the underlying Markovian %process. \hfill $\square$

\hspace{5mm} Note that this result implies that if evey combination of $c-2$ or more device failures are tolerable given the coding algorithm and allocation policy, $\xi_t$ has a simple closed form and thus exact expression can be calculated for $\phi_c(0)$ given by the recursive relation for $t= 1, \dots, c$
\begin{eqnarray}
\phi_t(0) = \prod_{i=0}^{t}\lambda_i + (\mu_{t-1} + \lambda_t)  \left[ \phi_{t-1}(0) - \prod_{i=0}^{t-1}\lambda_i  + \gamma_{t-1} \left( \prod_{i=0}^{t-2} (\gamma_i + \lambda_i) + \gamma_{t-3} \mu_{t-3} \prod_{i=0}^{t-4}\lambda_i \right)  \right]\label{Phip}
\end{eqnarray}
and the MTTDL expression become exact. Otherwise, our computations shall generate an upper bound for the actual MTTDL number (see why this is?).

\hspace{5mm} Beauty of these type of closed form expressions is that, one can deduce the reliability of large set of erasure coded arrays. For example, we can easily compute that 125 of RAID6 arrays with $m=8$ data devices (A total of $1PB$ user data stored on 1250 $1TB$ disk devices) has a durability number of $3$. This shows that parallel functioning conventional RAID schemes for data protection is quite unreliable as the scale of the stored data expands to PB ranges. If we reduce $m=8$ to $m=7$, allowing a total of $875TB$ user data storage in the same storage system increases the durability number to 6.

\subsection{Case Study: Generalized Model applied to Pyramid Codes}

\hspace{5mm} An interesting case study would be to apply the generalized Markov model of the previous section to one of the modern erasure codes such as Pyramid Codes of Microsoft Azure Storage \cite{Mic}. Pyramid codes are designed to improve the recovery performance for
small-scale device failures and have been implemented in archival storage \cite{Wildani}. Pyramid codes are constructed from standard MDS codes by constructing newer parity symbols from already existing parities in order to trade-off the \emph{recoverability}, \emph{coding overhead} and the \emph{average read overhead}, which are important parameters to optimize for a storage application. Let us use a $(16,12)$ MDS code as the basis for $(18,12)$ set of pyramid codes given in table \ref{Table41}. These recoverability and read overhead values are computed and presented in \cite{Mic}. First, we notice that recoverability values divided by $100$ gives us $q_i$ values for each one of the pyramid codes. The metric, \emph{average read overhead}, represents
the average number of extra device reads as an overhead in order to access each data block.   Let us consider an example of one block failure in the (18,12) MDS code to show how this metric is computed.
If the failure is a redundant block (6/18 chance), then the
data blocks can be accessed directly, so the average read
overhead is 1. Otherwise, the failure shall be a data block (12/18
chance), then the read overhead is twelve for the failed data block
and one for the rest of the eleven data blocks. Hence, the average read
overhead is (12+11)/12. Altogether, the average read overhead is
$1 \times 6/18+(12+11)/12 \times 12/18 \approx 1.61$.

\hspace{5mm}  In fact, we can find the average read overhead  for a generic $(n,k)$ MDS code when we have $j$ failures using the following generalized expression\footnote{This expression can easily be proved by Induction and thus the proof is omitted for space.}
\begin{eqnarray}
 \Phi_j = \sum_{i=0}^j \frac{(ik + k - i) \binom{n-k}{j-i}\binom{k}{i}}{k \binom{n}{j}}
\end{eqnarray}

\begin{table}
\footnotesize
\begin{center}
\begin{tabular}{ |l|l|l|l|l|l|l|l|l| }
\hline
\multicolumn{2}{ |c| }{Number of failed symbols/blocks} & 0 & 1 & 2 & 3 & 4 & 5 & 6 \\
\hline
\multirow{2}{*}{Generic MDS Code} & Recoverability ($\%$) & 100 & 100 & 100 & 100 & 100 & 100 & 100 \\
 & Avg. read overhead & 1.0 & 1.61 & 2.22 & 2.83 & 3.44 & 4.06 & 4.67 \\
\hline
\multirow{2}{*}{Basic Pyramid Code (BPC)} & Recoverability ($\%$) & 100 & 100 & 100 & 100 & 100 & 94.12 & 59.32 \\
 & Avg. read overhead & 1.0 & 1.28 & 1.56 & 1.99 & 2.59 & 3.29 & 3.83 \\
\hline
\multirow{2}{*}{Generalized Pyramid Code (GPC)} & Recoverability ($\%$) & 100 & 100 & 100 & 100 & 100 & 94.19 & 76.44 \\
 &  Avg. read overhead & 1.0 & 1.28 & 1.56 & 1.99 & 2.59 & 3.29 & 4.12 \\ \hline
\multirow{2}{*}{GPC w/o global symbols} & Recoverability ($\%$) & 100 & 100 & 100 & 100 & 97.94 & 88.57 & 65.63  \\
 &  Avg. read overhead & 1.0 & 1.28 & 1.56 & 1.87 & 2.32 & 2.93 & 3.85 \\
\hline
\end{tabular}
\end{center}
\caption{\footnotesize{A generic $(18,12)$ MDS code and few Pyramid codes with the associated recoverability/efficieny characteristics as given in \cite{Mic}}} \label{Table41}
\normalsize
\end{table}

\hspace{5mm} Although the average overhead is not the only metric effecting the repair process, for simplicity we assume repair rates to be inversely proportional to that metric. Let us define
\begin{eqnarray}
\overline{\mu}_j \triangleq  \delta  \mu \ln{( (j+1) \Phi_{j+1})}
\end{eqnarray}
where $\mu$ is the nominal rate of the repair per device and $\delta$ is a constant used to model the relative bandwidth constraint (with respect to an MDS code, for an MDS code it is normalized to $\delta = 1$) to reflect on the repair rates based on average read overhead metric. This formulation assumes an inverse exponential relationship between the nominal repair rate and the average read overhead. More field data is needed to confirm such a relationship.  

\hspace{5mm} Let $\chi_i$ be the average read overhead using one of the pyramid codes when $i$ devices fail. A closed form expression for $\chi_i$ for a given pyramid code with any desired level of hierarchy is not derived in the original paper and is an interesting open problem. We shall use the computed values of $\chi_i$ in \cite{Mic} for our MTTDL computation. Using the generalized Markov model of the previous section, let us further assume a homogenous repair strategy is used in the system i.e., $\mu_j = \frac{\overline{\mu}_j}{\ln{((j+1)\chi_{j+1}}))}$. Some results are shown in table \ref{Table42} using a nominal repair rate $\mu = 1/168$ (1 week mean repair time), $\eta = 10^{-3}$ and $\delta = 20$. We observe that basic and generalized pyramid codes provide better durability numbers thanks to their efficient repair mechanisms. As $\lambda$ gets close to zero, the frequency of repairs go down and hence the advantage of pyramid codes diminish. This can be observed with the MTTDL results given for $\lambda = 1/1.2M$. Another interesting observation is that given these computation parameter settings, global symbols are quite crucial for pyramid codes for maintaining a desired level of durability.

\begin{table}
\footnotesize
\begin{center}
\begin{tabular}{ |l|l|l|l|l|}
\hline
\multicolumn{2}{ |c| }{Failure/Nominal Repair rates} & $\lambda= \frac{1}{200K}$ & $\lambda= \frac{1}{500K}$ & $\lambda= \frac{1}{1.2M}$  \\
\hline
\multirow{2}{*}{Generic MDS Code} & MTTDL ($hours$) & 2.2e+15 & 6.4e+17 & 1.3e+20  \\
 & Durability ($nines$) & 11 & 13 & 15  \\
\hline
\multirow{2}{*}{Basic Pyramid Code (BPC)} & MTTDL ($hours$) & 1.3e+17 & 5.2e+18 & 1.7e+20  \\
 & Durability ($nines$) & 13 & 14 & 16 \\
\hline
\multirow{2}{*}{Generalized Pyramid Code (GPC)} & MTTDL ($hours$) & 1.32e+17 & 5.26e+18 & 1.76e+20  \\
 &  Durability ($nines$) & 13 & 14 & 16 \\ \hline
\multirow{2}{*}{GPC w/o global symbols} & MTTDL ($hours$) & 1.83e+14  & 3e+15 & 4.1e+16  \\
 &  Durability ($nines$)  & 10 & 11 & 12  \\
\hline
\end{tabular}
\end{center}
\caption{\footnotesize{MTTDL for various erasure codes.}} \label{Table42}
\normalsize
\end{table}

\section{Availability}

\hspace{5mm} One of the other important metrics of Reliability is \emph{availability}, defined as the fraction of time the system is able to provide access to the data through some redundancy scheme. Availability deals more with temporary inaccessibility when enough number of devices are unable to respond for the reconstruction of the data.  The lifetime of a device is made up of periods of availability and unavailability. Device \emph{uptime} means the time while the device is operational/online whereas the \emph{downtime} of the device refers to the period in which the device is unresponsive/offline. When the device is offline, the data it contains is temporarily unavailable and does not participate in the system unless it becomes online again. Devices can become unavailable for a variety of
reasons. For example, a storage device, node or networking
switch can be overloaded; the operating
system may crash or restart; the controller may experience
a hardware error; or the whole cluster of storage devices could be brought down for maintenance. The vast
majority of such unavailability events are transient and
do not result in permanent data loss \cite{Google}.

\begin{figure}[b!]
\centering
\includegraphics[angle=0, height=30mm, width=70mm]{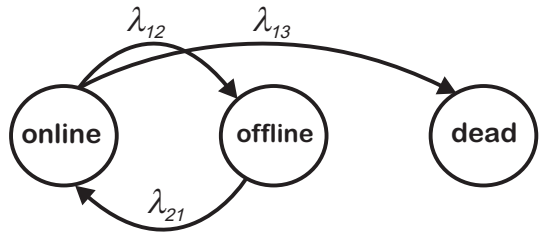}
\caption{Markov chain for availability of a device/node.}\label{fig:AvailMarkov}
\end{figure}

\hspace{5mm} One of the simplest definitions for availability can be given by the ratio of the expected uptime divided by the sum of the expected up and down times, i.e.,
\begin{eqnarray}
p_A \triangleq \frac{\mathbb{E} \emph{uptime}}{\mathbb{E} \emph{downtime} + \mathbb{E} \emph{uptime}}
\end{eqnarray}

\hspace{5mm} If we assume uptimes and downtimes to be exponentially distributed, the underlying random process becomes a Markovian, the state diagram of which can be shown as in Fig.  \ref{fig:AvailMarkov}. Let us say the mean uptime and downtime of a device are given by $t_{up}$ and $t_{down}$. Device availability is given by $p_{A} = t_{up}/(t_{up} + t_{down})$. Furthermore if the MTTF ($1/\lambda$) \footnote{Here we implicitly assumed that device lifetime is exponentially distributed. In fact it can be shown that this is the case if $\lambda t_{down} \ll 1$ \cite{Ramabhadran}. } of the device is sufficiently large with respect to mean uptime and down times ($\lambda t_{up} > \lambda t_{down} \gg 1$), we can show that \cite{Ramabhadran}, the transition rates of the continuous Markov model are
\begin{eqnarray}
\lambda_{12} = \frac{p_A - \lambda t_{up}}{t_{up}p_A},  \ \ \lambda_{13} = \frac{\lambda}{p_A}, \ \ \lambda_{21} = \frac{1}{t_{down}}.
\end{eqnarray}

 \hspace{5mm} Note that if the device never goes offline, i.e., $p_A = 1$ or $t_{down} = 0$, we have $\lambda_{13} = \lambda = 1/t_{up}$. Moreover, using the result of Lemma 1, we can conclude that the probability of a transition from state \textbf{online} to the state \textbf{dead} is given by
 \begin{eqnarray}
 p_{13} = \frac{\lambda_{13}}{\lambda_{12}+\lambda_{13}} = \frac{\lambda t_{up}}{p_A} = \lambda ( t_{up} +  t_{down})
\end{eqnarray}
which will be useful for next section's discussion.

 \subsection{Availability with Timeouts}

 \hspace{5mm} Although the state diagram is clear, the system cannot distinguish between a data/parity device that
is dead, and one that is merely offline. In both cases, the device is unresponsive and there is no way of knowing whether it is going to be back online again. This detection problem leads to an uncertainty about when to initiate the repair process. One of the mechanisms to
decide when to trigger a repair is to wait some period of time (called the \emph{timeout})
for the offline/dead replica to return to the online state. If the device returns online before the timeout
occurs, no repair is necessary; otherwise, the system assumes that it is dead, and initiates a repair process. Choosing a timeout period depends on how aggressively
the system wants to repair. The timeout period must be optimized for a given system to prevent from inefficient use of system resources (for example unnecessary repairs).

 \hspace{5mm} The following discussion is given in \cite{Ramabhadran} and will be restated here as it provides us useful information about the actual repair process with timeouts. They introduced the system parameter $\alpha$ such that the timeout period shall be given by $\alpha t_{down}$. The  parameter $\alpha$
allows the system to trade off how aggressively it responds to
the potential loss of a data/parity versus how many unnecessary
repair initiations are made.

 \hspace{5mm} Let $Y_{\alpha}$ denote the time that passes between the data/parity generation, and leave of the online state without return i.e., either the instant of irrecoverable error or a period of time exceeding the timeout. From the arguments of $\cite{Ramabhadran}$, we have the following lemma.

 \hspace{5mm} \textbf{Lemma 4 \cite{Ramabhadran}:} \emph{The expected value of the random variable $Y_\alpha$ is given by
\begin{eqnarray}
\mathbb{E}Y_\alpha = \frac{(1-p_{13})(1-e^{-\alpha})\left[t_{up} + t_{down}\left(1 - \frac{\alpha e^{-\alpha}}{1-e^{-\alpha}}\right)\right]}{p_{13} + (1-p_{13})e^{-\alpha}} + t_{up}
\end{eqnarray}
%\begin{eqnarray}
%\xi_t = \gamma_{t-3} \mu_{t-3} \prod_{i=0}^{t-4}\lambda_i
%\end{eqnarray}
}

\hspace{5mm} PROOF: Proof of this lemma is given in \cite{Ramabhadran}. \hfill $\square$

\begin{figure}[t!]
\centering
\includegraphics[angle=0, height=90mm, width=130mm]{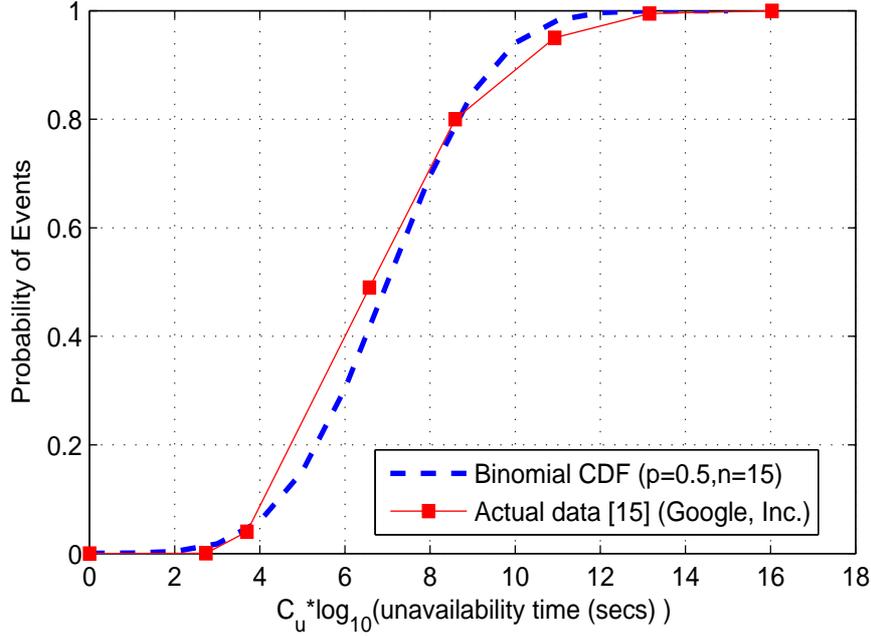}
\caption{Cumulative distribution function of the duration of node unavailability periods versus the binomial fit.}\label{fig:Google}
\end{figure}

 \hspace{5mm} A general practice of choosing an appropriate $\alpha$ can be given by solving the following equation given $\lambda, t_{up}$ and $t_{down}$
\begin{eqnarray}
%\frac{(1-p_{13})(1-e^{\alpha})\left[t_{up} + t_{down}\left(1 - \frac{\alpha e^{-\alpha}}{1-e^{-\alpha}}\right)\right]}{p_{13} + (1-p_{13})e^{-\alpha}} + t_{up} + \alpha t_{down} = 1/\lambda \\
%\frac{(1-p_{13})(1-e^{\alpha})\left[t_{up} + t_{down}\left(1 - \frac{\alpha e^{-\alpha}}{1-e^{-\alpha}}\right)\right]}{p_{13} + (1-p_{13})e^{-\alpha}} + t_{up} + t_{down} = 1/\lambda + (1-\alpha)t_{down} \\
\frac{\lambda (1-p_{13})(1-e^{-\alpha})\left[t_{up} + t_{down}\left(1 - \frac{\alpha e^{-\alpha}}{1-e^{-\alpha}}\right)\right]}{p_{13} + (1-p_{13})e^{-\alpha}} = 1+ \lambda (1-\alpha)t_{down} - p_{13} \label{eqn_avail}
\end{eqnarray}
where we can notice that solving this nonlinear equation is nothing but finding the point that satisfies $\mathbb{E}Y_\alpha + \alpha t_{down} = 1/\lambda$ i.e., expected time to timeout intersects with the linear line $y = 1/\lambda$. This in turn means the repair process is initiated and hopefully generates
exactly one new replica at every device lifetime. The same study also presents results regarding the insertion of memory to the overall repair process. Apparently, taking into account which devices or nodes are subject to repair, and when they return to online state, unnecessary redundancy can be eliminated to save the system resources. Interestingly though, when $\alpha$ is chosen meticulously it is shown that with-memory repair is not significantly better
than without-memory repair.

\subsection{Case Study: Google unavailability data}

\hspace{5mm} The data gathered  from tens of Google storage cells, each with 1000 to 7000 storage nodes, over a one year period \cite{Google} suggests that less
than 10\% of unavailability events last longer than 15 minutes. Let us express the unavailability event duration in units of seconds, and use a mapping of the duration metric using $C_u \times \log_{10}(``\textrm{unavailability event duration}")$, where $C_u$ is an appropriate constant. This change appropriately allows us to fit a binomial CDF to the real data as shown in Fig. \ref{fig:Google}. The binomial distribution has parameters $p=0.5$ and $n=15$ to match the data with $C_u = 3.7$. Although such is given for node unavailability in \cite{Google}, we expect similar trends with device/disk unavailabilities.

\hspace{5mm} Using the binomial distribution approximation based on the Google's data, $t_{down}$, the mean downtime can be found to be around $0.03$ hours. It is also reported in \cite{Google} that they typically wait 15 minutes ($0.25$ hours) before commencing recovery of data on unavailable nodes. This implies $\alpha = 8.\overline{3}$. We can either set an AFR i.e., using an exponential time distribution, $\lambda$ or the uptime $t_{up}$. It is usually more reasonable to measure  $t_{up}$ and through solving the equation (\ref{eqn_avail}), $\lambda$ can be calculated. With this value and the estimated repair rates, MTTDL computations can be performed as previously discussed.

\hspace{5mm} From Google's paper, it is convenient to assume an AFR value of $\% 4$ (also reported to be more realistic observed in many data centers and field measurements \cite{Google}) and calculate the $t_{up}$ as the standard uptime measurements are not reported in the paper. Using these assumptions, we solve the equation (\ref{eqn_avail}) for $t_{up}$ to be $218,089$ hours. This implies $\lambda_{13} = 0.99584$ which is close to one. Ofcourse, unavailability of devices can dramatically vary depending on the environment and correlation factor between other functional system components. True values shall change our calculations but definitely not the methodology presented here.

\section{Multi-Dimensional Markov Models for Cold Storage}

\hspace{5mm} In this section we focus on a specific data storage scenario which heavily applies to archival scenarios where the data is written once and read multiple times and rarely. In addition, with this application in mind, we disclose important components of a cold storage system in which both reliability and availability need to be considered concurrently. Therefore, this is a solid use case where the foundations that we laid out in previous chapters about the reliability and availability. The contents of this section is largely covered in \cite{sarslan2020}.

\hspace{5mm} Cold data storage systems are  used to allow long term digital preservation for institutions' archive.  The common functionality among cold and warm/hot data storage is that the data is stored on some physical medium for read-back at a later time. However in cold storage, write and read operations are not necessarily done in the same exact geographical location. Hence, a third party assistance is typically utilized to bring together the medium and the drive. On the other hand, the reliability modeling of such a decomposed system poses few challenges that do not necessarily exist in other warm/hot storage alternatives such as fault detection and absence of the carrier, all totaling up to the data unavailability issues. In this paper, we propose a generalized non-homogenous Markov model that encompasses the aging of the carriers in order to address the requirements of today's cold data storage systems in which the data is encoded and spread across multiple nodes for the long-term data retention. We have derived useful lower/upper bounds on the overall system availability. Furthermore, the collected field data is used to estimate parameters of a Weibull distribution to accurately predict the lifetime of the carriers in an example scale-out setting.

\hspace{5mm} In an accurate reliability modeling for cold data storage systems (such as tape), the media failure, faulty drive behaviour, failure detection, data spreading -- that is how much the data is spread across different independent cold storage nodes -- should all be taken into account \cite{arslan2014mds}. Particularly, the detection of faulty system behaviour and  lost data can be done by periodic system/data check operations which is known as \textit{scrubbing} in disk-based systems \cite{oprea2010clean}. Such a scrubbing process can be incorporated into reliability models \cite{ryu2009effects,schwarz2004disk,iliadis2008disk}. In contrast, in case of carrier failures, data is not lost but would be temporarily unavailable until the carrier is replaced/fixed by the maintenance team. Therefore, the actual reliability model should treat data durability and availability individually and be able to neatly merge these important data-related performance indicators.

\hspace{5mm} With regard to cold storage systems, several past research studied application-specific back-up systems in which keeping the copies of the data (replication) was the primary means of providing durability \cite{constantopoulos2005}. Although they have paid attention to failure detection problem, they did not configure their model for true cold storage environment requirements i.e., the necessity of carrying media from one location to another, detrimental effects of mechanical components, unavailability of carriers and drive-related hard errors. In \cite{HanChan}, the work is extended to cover 4-copy case where the backup system consists of both tapes and hard drives with different failure and repair rates i.e., a heterogeneous storage network. Since the storage media and internal mechanics are different for tape systems and hard drives, the proposed model quickly gets complicated as the number of copies increase. Therefore, extending it beyond 4-copy seems to be quite challenging and no systematic extension is proposed in the same line of work. In  \cite{ivanichkina2015}, a two dimensional Markov process is proposed for modeling explicit and latent errors in disk-based distributed storage systems in which failure detection is assumed to take almost no time. Since only disk-based systems are considered, the study is not extended to take into account the presence of carriers. 

\hspace{5mm} Due to the complexities of data dependent density estimation and the time dependent carrier aging phenomenon, a multi-dimensional Markov model is needed and can be implemented as a simulation platform to estimate the distributions of time to data-loss and data-unavailability. Known statistics such as MTTD and MTTDU can be derived from the simulation data and compared for various choices of system parameters as well as few theoretical results known for a limited parameter space. 

\subsection{General Model Description}

\hspace{5mm}  The proposed reliability model incorporates data--driven density estimation (a known distribution fit) and a continuous time Markov process to accurately estimate the density of the system data loss, and associated first order statistics such as MTTDL and MTTDU metrics. The model has two types of states: \textit{node} states and \textit{system} states all combined and characterized by the Markov states. Although the proposed model is generally applicable to any type of cold storage, we specifically consider a scale--out tape library system.

\hspace{5mm} Although we treat the number of node states to be any arbitrary number, for simplicity, we will give our examples by proposing three different  node states that are the most commonly assumed in cold data storage community. There are defined to be \textit{Available} (A), \textit{Failed} (F) and \textit{Detected} (D) for a given node and we automatically generate the associated Markov states of the overall system. We would like to remind that typical continuous Markov processes are heavily used  for warm/hot storage devices  consisting of hard disk or solid state device arrays (such as given in \cite{Hafner2006}). Previous work focused on Markov processes with typically two node states, namely Available and Failure. Indeed, our treatment makes our model a generalized version of all the previous Markov models used with configurable parameters.  Although it becomes impossible to derive closed form expressions, we shall use approximations to derive analytical results for performance metrics such as lower and upper bounds on the mean statistics.

\hspace{5mm} To illustrate one instantiation of the proposed model, let us assume that the cold storage system is protected by a $(\Tilde{n},\Tilde{k})$ Maximum Distance Separable (MDS) erasure code where $\Tilde{n}$ is the codeword (block) and $\Tilde{k}$ is the payload lengths, respectively. Note that with this setting, the conventional replication (copy) system corresponds to $\Tilde{k}=1$. The quantity $r = \Tilde{k}/\Tilde{n} = k/n$ is termed as the rate of the code. In other words, since we use MDS codes, we assume $\Tilde{n}$ and $\Tilde{k}$ to be a multiple of $n$ and $k$, respectively. Due to encode/decode complexity and without loss of generality, we shall assume $n = \Tilde{n}$ and $k = \Tilde{k}$ throughout the document to convey the main idea.

\hspace{5mm}  We realize that node states and Markov states are not the same, in fact, a Markov state can have triple node states. We can visualize each Markov state consisting of three buckets counting the number of nodes having each of A, F and D node states. Furthermore, we assume nodes are exactly the same type and fail with the same rate $\lambda$ i.e., homogeneous storage network. This means that due to a physical and irreversible error, data cannot be read from the tape that resides in that node. In addition, we have three more processes running in the system; two of them are the concurrent and identical data and carrier repairer processes and the third one is the concurrent and identical failure detector process. In this study, we assume that  an error detection process is run on tapes and repair them whenever an error is detected. In addition, media carriers such as robots can also be repaired since without their availability, all data operations will cease. Robot and data repairer as well as failure detector processes are assumed to be exponentially distributed with rates $\phi$, $\mu$ and $\theta$, respectively.

\hspace{5mm} The complexity of the proposed continuous Markov process is strongly tied to the total number of system states which are a function of node states. For $s > 1$, i.e., number of node states being greater than one\footnote{In our formulation, we assume that node states A and F are naturally present in any reliability  model. In further extensions of the model proposed in this study, different processes can be incorporated such as detection, participation and aging.}, suppose we have $i$ available nodes with $k \leq i \leq n$ (each containing a single data chunk - remember this is a requirement for perfect data reconstruction) then we shall have $s-1$ node states to share a total of $n-i$ data chunks. In this case, the total number of decompositions is given by 
\begin{eqnarray}
\binom{n-i+s-2}{s-2}
\end{eqnarray}

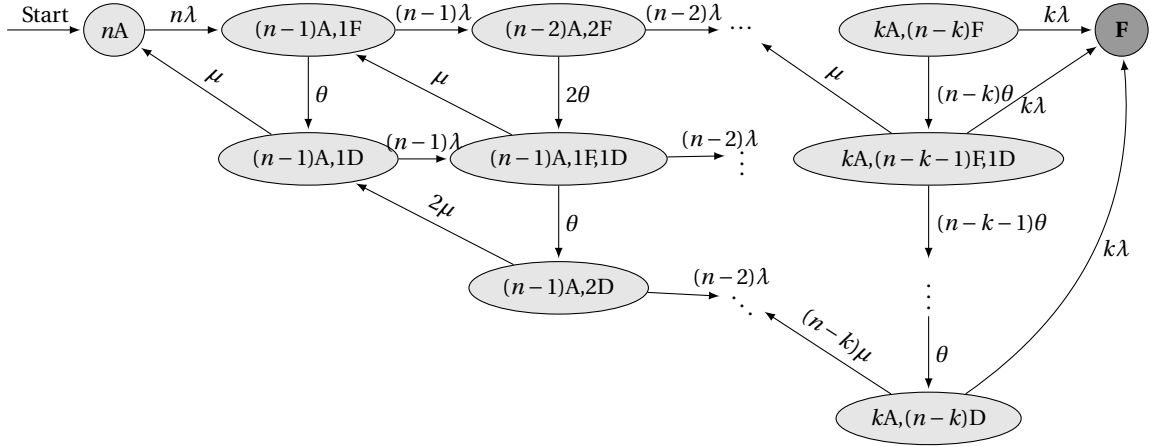
\begin{figure}
\footnotesize
\centering
%\hspace*{-2cm}
\begin{tikzpicture}
        % Setup the style for the states
        \tikzset{node style/.style={state, fill=gray!20!white}}
        \tikzset{node style2/.style={state, fill=gray!80!white}}
    
        \node[node style, ellipse]               (I)   {$n$A};
        \node[node style, ellipse, right=of I]   (II)  {$(n-1)$A,$1$F};
        \node[node style, ellipse, below=of II]  (IIb)  {$(n-1)$A,$1$D};
        \node[node style, ellipse, right=of II]  (III) {$(n-2)$A,$2$F};
        \node[node style, ellipse, below=of III]  (IIIb)  {$(n-1)$A,$1$F,$1$D};
        \node[node style, ellipse, below=of IIIb]  (IIIbb)  {$(n-1)$A,$2$D};
        \node[draw=none,  right=of III]   (IIIn) {$\cdots$};
        \node[draw=none,  ellipse, below=of IIIn]   (IIInb) {$\vdots$};
        \node[draw=none,  below=of IIInb]   (IIInbb) {$\ddots$};
        \node[node style, ellipse, right=of IIIn] (n)   {$k$A,$(n-k)$F};
        \node[node style, ellipse, below=of n]  (nb)  {$k$A,$(n-k-1)$F,$1$D};
        \node[draw=none, below=of nb]  (n-e) {$\vdots$}; 
        \node[node style, ellipse, below=of n-e] (neb) {$k$A,$(n-k)$D};
        \node[node style2, right=of n] (FF)   {\textbf{F}};
        %\node[draw=none, right=of n] (Failure)   {};
        
        % Draw empty nodes so we can connect them with arrows
        \node[draw=none, left=of I]   (a)    {};
    
    \draw[>=latex, auto=left, every loop]
         (a)   edge node {Start}                         (I)
         (II)  edge node {$\theta$}        (IIb)
         (IIb) edge node[sloped, above] {$\mu$} (I)
         (III) edge node {$2\theta$}       (IIIb)
         (IIb) edge node {$(n-1)\lambda$} (IIIb)
         (IIIb) edge node {$\theta$}       (IIIbb)
         (IIIb) edge node[sloped, above] {$\mu$} (II)
         (IIIb) edge node [very near end] {$(n-2)\lambda$} (IIInb)
         (IIIbb) edge node[sloped, above] {$2\mu$} (IIb)
         (IIIbb) edge node {$(n-2)\lambda$} (IIInbb)
         (n)   edge node {$(n-k)\theta$}                    (nb)
         (nb) edge node[sloped, above] {$\mu$}                    (IIIn)
         (nb) edge node[below] {$k\lambda$}                    (FF)
         (nb)   edge node {$(n-k-1)\theta$}                    (n-e)
         (n-e)   edge node {$\theta$}                    (neb)
         (neb) edge node[sloped, above] {$(n-k)\mu$}                    (IIInbb)
         (neb) edge[above, bend right, right=0.1] node {$k\lambda$} (FF)
         %(nb) edge node {$\lambda$}                (FF); 
         (I)   edge node {$n\lambda$}   (II)
         (II)  edge node {$(n-1)\lambda$}  (III)
         (III) edge node {$(n-2)\lambda$} (IIIn)
         (n) edge node {$k\lambda$}                (FF); 
         %
         %(n) edge node {$3t5t5t$}               (Failure);
         
    \end{tikzpicture}
    \caption{Generalized Markov model for $s = 3$ node states i.e., represented by Availability (A), Detection (D) and Failure (F). The state $\textbf{F}$ denotes the total failure/unavailability. Note that since $s=3$, corresponding Markov model can be represented on a two dimensional plane. In general, Markov model is $s-1$ dimensional.}
    \label{GMMs3}
\end{figure}

\hspace{5mm}  For instance, Fig. \ref{GMMs3} shows all of the system states (Markov states) as a function of $k$ and $n$ if the node states are $s=3$, represented by A, D and F. Yet in another case, the state ``Queued for Service: QS" can be added to make number of node states $4$, i.e., $s=4$. The closed form expression to calculate the total number of Markov states (for general $s$) is given by (including the total failure state)
\begin{align}
N_s &= \sum_{i=k}^n \binom{n-i+s-2}{s-2} + 1 = \sum_{i=0}^{n-k} \binom{i+s-2}{s-2} + 1  \nonumber \\ &= \binom{n-k+s-1}{n-k} + 1 \label{nseqn}
\end{align}

\hspace{5mm}  Note that for the general case, we can further express $N_s$ as follows
\begin{eqnarray}
N_s &=& 1 + \binom{n-k+s-1}{n-k}  \label{eqn12}\\
&=& 1 + \binom{n-k+1}{1} + \sum_{i=2}^{s-1} \binom{n-k+i-1}{i} \label{eqn13}\\
&\geq& \sum_{i=0}^{s-1} \binom{n-k+1}{i}  \label{eqn14}
\end{eqnarray}
where we clearly see that equality in equation \eqref{eqn14} holds only for $s=2,3$. For small $s$,  equation \eqref{eqn14} can be used as an accurate approximation. Note that going from  \eqref{eqn12} to \eqref{eqn13}, we have used induction. In the following, we provide  tighter upper and lower bounds for $N_s$ and asymptotically analyze the complexity of the final Markov reliability model.

\hspace{5mm} Observe that using Vandermonde convolution for the expression given for $N_s$, we can rewrite for it $s > 1$
\begin{align}
N_s &=  \binom{n-k+s-1}{s-1} + 1  \\
&=  \sum_{j=1}^{s-2} \binom{s-2}{j} \binom{n-k+1}{j+1} + n - k + 2 \\
&\geq \sum_{j=2}^{s-1} \left(\frac{s-2}{j-1}\right)^{j-1} \binom{n-k+1}{j} + \sum_{j=0}^{1} \binom{n-k+1}{j} \label{eqn50}
\end{align}
from which we can deduce that
\begin{eqnarray}
N_s \geq \sum_{j=0}^{s-1} \binom{n-k+1}{j}. \label{eqn51}
\end{eqnarray}
Note that the lower bound in Eq. \eqref{eqn50} is a tighter compared to one in \eqref{eqn51}. For the upper bound, we observe that
\begin{eqnarray}
N_s &=& \sum_{j=1}^{s-2} \binom{s-2}{j} \binom{n-k+1}{j+1} + n - k + 2 \nonumber \\
&=&  \sum_{j=2}^{s-1} \binom{s-2}{j-1} \binom{n-k+1}{j} + \sum_{j=0}^{1} \binom{n-k+1}{j} \label{eqn49} \nonumber \\
&\leq&  \sum_{j=0}^{s-1} \binom{s-1}{j} \binom{n-k+1}{j} = \sum_{j=0}^{s-1} \binom{s-1}{j} \binom{n-k+1}{n-k+1-j} \label{eqn50a} \nonumber \\
&=& \binom{s+n-k}{s-1}
\end{eqnarray}
where \eqref{eqn50a} results from \eqref{eqn49} using Pascal's triangle inequality which for any positive $m \geq c$ is given by
\begin{eqnarray}
\binom{m}{c} = \binom{m-1}{c-1} + \binom{m-1}{c}
\end{eqnarray}

Since $\binom{s+n-k}{s-1} = \binom{s+n-k}{n-k+1}$ we can deduce that for a fixed $s \ll n$ and a scaling  $k$ that is linear in some large $n$, i.e., $k=\alpha n$ for any $\{\alpha: 0 < \alpha < 1\}$, the total number of system (Markov) states will scale with $O(n^{\min\{s-1,n-k\}}) = O(n^{\min\{s-1,(1-\alpha)n\}})$ = $O(n^{s-1})$. In other words, the complexity of our simulation framework (and the corresponding Markov chain) grows exponentially in the number of node states unless the rate of the code goes to unity ($r \rightarrow 1$), i.e., $k$ becomes sublinear in $n$ with constant $n-k < s$. In that case the total number of system states would scale with $O(n^{n-k})$ with no dependence on the number of node states, $s$.

\subsection{A special case: 3-node-state Markov Model}

\begin{table}[b!]
\normalsize
\centering
\begin{tabular}{|l|l|l|}
\hline \hline
Curr. State & Destinations     & Transition Rate \\ \hline
$i$A,$j$F,$z$D    & $(i-1)$A,$(j+1)$F,$z$D & $i\lambda$               \\ \hline
$i$A,$j$F,$z$D     & $i$A,$(j-1)$F,$(z+1)$D & $j\theta$               \\ \hline
$i$A,$j$F,$z$D     & $(i+1)$A,$j$F,$(z-1)$D & $z\mu$               \\ \hline \hline
\end{tabular}
\caption{State transitions and rates}
\label{statestable}
\end{table}

\hspace{5mm}  As introduced earlier, let us assume we have three node states: A, D and F. Therefore, there are only three (in general $s$) destinations that a next state change could result  in. For instance, for a given state index $(i$A,$j$D,$z$F$)$, Table \ref{statestable} summarizes all the state indexes as possible destinations. In the table, DS stands for destination state and $i,j,z$ should satisfy the inequalities
\begin{eqnarray}
k \leq i \leq n,  \ 
0 \leq j \leq n - k, \
0 \leq z \leq n - k,  \ 
i + j + z = n \nonumber
\end{eqnarray}

\hspace{5mm}  One of the things that is not accounted for in the simulation model is a transformation method from the three-index state name to a single index name that runs between 0 and $N_3-1$ for simulation convenience. One straightforward method is to let the common index $c_{ind}$ to be 
\begin{eqnarray}
c_{ind} &=& \frac{1}{2}(j+z)(j+z+1) + z + 1 \nonumber\\
&=& \frac{(n-i+1)!}{2(n-i-1)!} + z + 1 
= \binom{n-i+1}{2} + z + 1
\label{eqn3}
\end{eqnarray}
where $j+z = n-i$ is replaced to find the first equality. Note that there is a one-to-one relationship between $c_{ind}$ and $(i,j,z)$ and  we can similarly find the inverse transform of the index $c_{ind}$. To find ($i,j,z$) from a given $c_{ind}$, we first need to find the maximum $i \in \{k, k+1, \dots, n\}$ such that $c_{ind} > i(i+1)/2$. From equation \eqref{eqn3}, we can find $z$ since we know $n$ and $i$. Finally, we use the fact that $i+j+z = n$ to determine $j$. 

\hspace{5mm}  Also the maximum of $c_{ind}$ shall be achieved with the state index $(k,0,n-k)$. Since $N_s = \max\{c_{ind}\} + 1$, the following can be shown to be true
\begin{eqnarray}
N_3  &=&  \binom{n-k+1}{2} + n-k + 2 \\
&=& \frac{1}{2}
\left((n-k+1)(n-k) + 2(n-k+2)\right) \\
&=& \frac{1}{2}(n-k+2)(n-k+1) + 1 \\
&=& \sum_{i=1}^{n-k+1}i + 1 = \sum_{i=0}^{n-k}(i + 1) + 1
\end{eqnarray}

\hspace{5mm}  Note that the final equality is the same as the result given by the Eqn. \eqref{nseqn} when $s = 3$. In order to derive performance expressions and apply the general model to a particular practical application, we shall assume $s=3$ for the rest of our discussions.

\subsection{Failure Types and Carrier Unavailability in Cold Storage}

\hspace{5mm}   Hard error scenarios are well understood in warm/hot storage realms i.e., when the drive and the storage medium are tightly coupled. In addition, there is no separate carrier availability problem due to this coupling. However, modeling and incorporating the hard errors as well as the carrier availability all at the same time into a reliability model is much more challenging in a cold data storage context.

\hspace{5mm}  Let us consider tape library systems as an example use case. One of the fundamental challenge is that the robots (carrier devices) can make some given number of exchanges--swaps (round-trips) before failure. Since such a constraint depends on time and the frequency of use (load of the system), this would add non-homogeneity to the Markov model at hand. Also, there are two driving forces for aging in the same system: (1) The user data access pattern which is usually less dominant in a cold storage setting and (2) the internally generated access requests due to system/data repair operations which will lead to extra robot exchanges, drive load/unload cycles, tape positioning etc., to be able to meet the system reliability goals. Similar observations can be made for other popular cold storage alternatives. 

\hspace{5mm} For simplicity, we have constrained our set and assumed three dominant factors two of which directly affects the data durability while the other only changes the data unavailability. One of them is the drive read failures. This type of hard error is also pretty prevalent in warm/hot storage where drives become unable to read the data due to an uncorrectable error by the virtue of internal error correction decoding failure of the drive. Uncorrectable error rate (UCER) is usually given in terms of errors per number of bytes or bits read and are usually due to random noise effects. This hard error mechanism is usually assumed to be time-independent and seeing at least one read error can be calculated by independence assumption, given by
\begin{eqnarray}
\epsilon = 1 - (1 - UCER)^{tape\_capacity}
\end{eqnarray}
where we assume the worst case scenario i.e., bulk reads i.e., the entire tape is read and $tape\_capacity$ is the number of bytes a tape can store. Note that in cold storage, this worst case scenario is quite common. 

\hspace{5mm} The other is the storage medium (tape) damage. In our work, we model the probability of tape damage due to manufacturing reasons (in the infant mortality period) at the onset or external factors such as humidity, pressure and stringent temperature conditions later on in their lifetime. Such factors are assumed to be static and persists after failure detection and correction throughout the lifespan of the data stored in the cold storage. For this rationale, we used $\kappa > 0$ to model this damage probability. More specifically, we refer to tape damages to be $\kappa \times 100$ percent of all the tapes contained in a given library.

\hspace{5mm} Suppose that library robots (carriers) can make $m$ number of exchanges (round-trips) before they fail and eventually become unable to complete tasks initiated by the libraries including the detection and repair processes. We assume the robot types and qualities are identical in all of the libraries. Based on the available data extracted from our local library systems, we shall show that $m$ can be modelled as a  Weibull distributed random variable with some shape($g$) and scale($y$) parameters. 

\hspace{5mm} It is typical to assume the time between exchanges to be exponentially distributed  with rate $\omega$. The rate $\omega$ depends on a number of parameters such as the number of users using the system, the total number of libraries in a scale-out setting, the time of the year etc. The time to robot failure (characterized by the random variable $Y$) is therefore the sum of $m$ exponential distributions each with the same rate $\omega$ i.e., Gamma distributed with the following pdf
\begin{align}
f_{Y}(y; \omega, m) = \frac{\omega^m y^{m-1}e^{-\omega y}}{\Gamma(m)} = \frac{\omega^m y^{m-1}e^{-\omega y}}{(m-1)!} \ \ \omega,m > 0.
\end{align}
where $\Gamma(.)$ is the complete Gamma function. Note that since Gamma distribution is NOT memoryless, it requires aging to be taken care of by inserting the time dependent conditional CDF (carrier survival probabilities) given by
\begin{eqnarray}
\beta(t) &=& P(Y > t | l \textrm{ exchanges made}) \nonumber \\ &=& 1 - P(Y < t | l \textrm{ exchanges made}) \nonumber \\ 
&=& \frac{\Gamma(l, \omega t)}{\Gamma(l)} = \frac{\Gamma(l, \omega t)}{(l-1)!}, \ \ \ t > 0.
\end{eqnarray}
where $l < m$ exchanges are assumed to be made by the robot. In that case, the latter conditional probability is also Gamma distributed with the pdf $f_Y(y; \omega, l)$. Also, the included upper incomplete Gamma function is given by
\begin{eqnarray}
\Gamma(l, \omega t) = \int_{\omega t}^{\infty} x^{l-1} e^{-x} dx 
\end{eqnarray}

\hspace{5mm} On the other hand, the hard error rate is given by $\eta = 1 - (1-\epsilon)(1-\kappa)$. Note that we need to modify the model to compensate for the hard errors.  Note that hard errors split the state transition from the current state $(i$A,$j$F,$z$D$)$ to the destination state $((i-1)$A,$(j+1)$F,$z$D$)$ which originally happens with rate $i\lambda$. Similar to the observations in previous studies, for each state with $i > k$, we need a transition to the total failure ($\textbf{F}$) state to be able to incorporate the hard errors.  With $i$ available nodes, we can tolerate up to $i - k - 1$ concurrent hard errors to successfully make it to the state $((i-1)$A,$(j+1)$F,$z$D$)$. Assuming independence, this happens with probability
\begin{eqnarray}
\Delta_i := \sum_{l=0}^{i-k-1}  \binom{i}{l}\eta^l(1-\eta)^{i-l} 
= 1 - I_{\eta}(i-k, k+1)
\end{eqnarray}
where $l$ is the number of hard errors that occur at the same time while rebuilding or during regular data checks. Also we have used the regularized beta function $I_x(a,b) = B(x; a, b)/B(a,b)$ instead to avoid the instability and precision issues of the binomial CDF. Here $B(x; a,b)$ is called the incomplete beta function and is given by
\begin{eqnarray}
B(x; a,b) = \int_{0}^x v^{a-1} (1-v)^{b-1}dv
\end{eqnarray}
and its complete version $B(a,b) = B(1; a,b)= \frac{\Gamma(a)\Gamma(b)}{\Gamma(a+b)}$. Note that we have the following limit
$\lim_{\eta \rightarrow 1} \Delta_i = 0 $
where the proposed model reduces to a simple transition from ($n$A,$0$F,$0$D) to $\textbf{F}$ with rate $n \lambda$. Since the hold times are assumed to be exponential, the mean time to stay in that state is $1/n\lambda$ which can be thought as the lower bound on the durability of the system. On the other hand, for small $\kappa$ and $\epsilon$, the upper bound can closely be approximated by the reliability of the system presented in Fig. \ref{GMMs3}. Finally, we summarize the new state transition table in Table \ref{dest_table2} with indexes satisfying
\begin{eqnarray}
k + 1 \leq i \leq n,  \ \ \
0 \leq j \leq n - k -1 \nonumber \\ 
0 \leq z \leq n - k - 1,  \ \ \
i + j + z = n. \nonumber
\end{eqnarray}

\hspace{5mm} Note that if we redraw the overall Markov system given in Fig. \ref{GMMs3} to incorporate hard errors,  it will make it look more complicated. To reach such a transition table, we have made a few assumptions that can be listed as follows.
\begin{itemize}
    \item In a typical repair process, only $k$ tapes are selected for repair process. In case of locally repairable codes \cite{LRC}, this number can be reduced. Alternatively, more than $k$ tapes can be requested and only earliest $k$ reads can be used to improve performance. 
    \item Hard errors are assumed to be independent i.e., data cannot have more than one segment(data chunk)  in the same library node.
\end{itemize}

\begin{table}[t!]
\normalsize
\centering
\begin{tabular}{|l|l|l|}
\hline \hline
Curr. State & Destinations     & Transition Rate \\ \hline
$i$A,$j$F,$z$D    & $(i-1)$A,$(j+1)$F,$z$D & $i\lambda \Delta_i$               \\ \hline
$i$A,$j$F,$z$D    & \textbf{F} & $i\lambda(1-\Delta_i)$             \\ \hline
$i$A,$j$F,$z$D     & $i$A,$(j-1)$F,$(z+1)$D & $j\theta$               \\ \hline
$i$A,$j$F,$z$D     & $(i+1)$A,$j$F,$(z-1)$D & $z\mu$               \\ \hline \hline
\end{tabular}
\caption{State transitions and corresponding transition rates}
\label{dest_table2}
\end{table}

\begin{center}
\begin{figure*}[htp!]
\footnotesize{
 \[
 \textbf{Q} =  \left[ \begin{array}{ccccccc}
-n\lambda   & n\lambda\Delta_n & 0 & 0 & \dots & 0 & n\lambda(1-\Delta_n) \\
0 & -(\theta + (n-1)\lambda) & \theta & (n-1)\lambda\Delta_{n-1} &\dots & 0 & (n-1)\lambda(1-\Delta_{n-1}) \\
\mu & 0 & -(\mu + (n-1)\lambda) & 0 & \dots & 0 & (n-1)\lambda(1-\Delta_{n-1}) \\
0 & 0  & 0 & 0 & \dots & 0 & (n-2)\lambda(1-\Delta_{n-2}) \\
\vdots & \vdots & \vdots &  & \ddots & \vdots & \vdots \\
0 & 0 & 0 & \dots &   & -(k\lambda+(n-k)\mu) & k\lambda \\
0 & 0 & 0 & \dots & 0 & 0 & 0 \end{array} \right]\]
%\caption{Transition Rate Matrix given for the Markov model shown in Fig. \ref{GMMs4}.}
%\label{rate_matrix}
}
\end{figure*}
\end{center}

\subsection{Transition Rate, Probability Matrices and Carrier Availability}

\hspace{5mm} The transition rate matrix (TRM) $\textbf{Q}$ (as shown above) is a $N_s \times N_s$ real valued matrix, whose entries $q_{ij} \geq 0$ for $i \not= j$ represent the rate departing from state $i$ and arriving in state $j$. The transition rate matrix for our generalized model is shown below. Note that we have included the total failure state $(\textbf{F})$ as part of the matrix and hence the last row becomes all-zero vector. We notice that the diagonal entries satisfy
\begin{eqnarray}
q_{ii} = - \sum_{j \not= i} q_{ij} \Rightarrow  \sum_{j} q_{ij} = 0 \ \ \forall i\in\{0,1,\dots,N_s-1\}
\end{eqnarray}
which means the rows of the matrix must sum to zero. 

\hspace{5mm} Furthermore, let us define $\overline{\textbf{Q}}$ in which entries are defined as $\overline{q}_{ij} := q_{ij}/|q_{ii}|$ for all $i$ and $j$. Then, the transition probability matrix (TPM) is given by $\textbf{P} = \textbf{I} + \overline{\textbf{Q}}$. Note that this is the precise version of \textit{uniformization} technique\footnote{In that uniformization technique, $q_{ii}$ is replaced with $\gamma \geq \max |q_{ii}|$.} that compute transient solutions of finite state continuous-time Markov chains, by approximating the process using a discrete time Markov chain. This formulation will be useful when we derive the upper bound on the performance. 

\hspace{5mm} The treatment of the previous subsection did not include the availability of the carriers in the state transition matrix. One of the observations is that although the node failures (e.g. tape failures) are independent of robots' availability, node repair and failure detection mechanisms are highly dependent on the availability of carriers (robots) i.e., the rates that describe failure detection and node repair must be time-dependent as well. As the time passes by, detection and repair rates will go down unless carriers are updated sufficiently fast. 

\hspace{5mm} When nodes fail due to various reasons, failure detection process  immediately commences. Similarly, when these failures are detected, the associated repair process starts immediately. So for a given operating time $t$, system robots will not be of the same age and quality (due to potential replacements etc). This leads to unequal treatment of storage nodes and our simulation setup must keep track of indexes for which robots are replaced in order to  model the aging phenomenon. 

\subsubsection{Time-dependent Failure Detection}

cFor simplicity, let us assume each node has a single carrier (through averaging arguments, it can be generalized to multiple carriers without changing the following discussion) and let $\psi_o(t)$ be the probability of $o  \in \{0,1,\dots,i\}$ available robots (conditioned on a specific set of $i$ nodes) in the system at time $t$ with survival probabilities $\beta_{s_1},\beta_{s_2}, \dots, \beta_{s_i}$ where $s_m \in \{1,2,\dots,n\}$. It can be shown that 
\begin{eqnarray}
\psi_o(t) = \Re\{F_{\mathcal{P}}(o)\} - \Re\{F_{\mathcal{P}}(o-1)\}
\end{eqnarray}
where $\Re\{.\}$ denotes the real part and $F_{\mathcal{P}}(.)$ is the  CDF of Poisson binomial distribution given by
\begin{align}
F_{\mathcal{P}}(o) = \frac{1}{n+1} \sum_{l=0}^i e^{-\frac{2\sqrt{-1}\pi l o}{n+1}} \prod_{m=1}^{i} (1 + (e^{\frac{2\sqrt{-1}\pi l}{n+1}}-1)\beta_{s_m}(t))
\end{align}
where $\sqrt{-1}$ is the complex number that is a solution of the equation $x^2 = -1$. On the other hand, since in our study we assume failures, detections and repairs all to be exponentially distributed and detection and carrier repairs can only happen consecutively, the natural consequence of sum of multiple independent exponential distributions is no surprise. However, to be able to make our later analysis analytically tractable, we will use a first order approximation in this subsection\footnote{Although in numerical result section, we will show that this assumption is a good approximation by simulating the actual distributions.}. More specifically, we will assume the sum of $x+1$ exponential distributions with rates $\phi,\theta_1,\dots,\theta_x$ to be approximately exponentially distributed with rate $R_{\boldsymbol{\theta}}(\phi)$ given by
\begin{eqnarray}
R_{\boldsymbol{\theta}}(\phi) = \frac{1}{\frac{1}{\phi} + \sum_{c=1}^x \frac{1}{\theta_c}} \label{exptailapprox1}
\end{eqnarray}
where $\boldsymbol{\theta} = [\theta_1, \dots, \theta_x]$. When a failure event is detected by the system, a state transition happens from the originator state $(i$A,$j$F,$z$D$)$ to the destination state $(i$A,$(j-1)$F,$(z+1)$D$)$ for $j>0$. While performing the detection, we need $j$ robots to complete the process, if found less, say $b < j$, then we need to repair $j-b$ robots to have a total of $j$ robots to work on the detection process. Suppose that at time $t$, we condition on having $l$ failed robots satisfying $0 \leq l \leq j \leq n - k$. Then the conditional repair rate i.e., the rate of making the detection transition in the Markov model is given by
$(j-l)\theta + lR_{\boldsymbol{\theta}}(\phi)$. Thus, summing over all possibilities of $l$, we get the unconditional node failure detection rate given by
\begin{eqnarray}
\theta_j(t;\phi) &=& \sum_{l=0}^j \left((j-l)\theta + lR_{\boldsymbol{\theta}}(\phi)\right) \psi_{j-l}(t) \\
&=& j\theta - (\theta - R_{\boldsymbol{\theta}}(\phi))\sum_{l=0}^j l\psi_{j-l}(t)  \\
&=& j\theta - (\theta - R_{\boldsymbol{\theta}}(\phi))\sum_{m=1}^j (1-\beta_{s_m}(t)) \\
&=& j\theta - \frac{\theta^2}{\theta + \phi} \sum_{m=1}^{j} (1-\beta_{s_m}(t))
\end{eqnarray}
where $s_m \in \{1,\dots,n\}$. Notice that we have the inequality for any $t$,
\begin{eqnarray}
j\theta - \frac{j\theta^2}{\theta+\phi} \leq \theta_j(t;\phi) \leq j\theta
\end{eqnarray}
which implies that as $\phi \rightarrow \infty$ i.e., robot repairs being instantaneous, the detection rate would be $j\theta$ which is the same as that of without any robot failures as given in Fig. \ref{GMMs3}. 

\subsubsection{Time-dependent Carrier Repair} 
\hspace{5mm}  After a node failure is detected, our system immediately begins the repair process and the completion of the repair process implies a state transition  from the originator state $(i$A,$j$F,$z$D$)$ to the destination state $((i+1)$A,$j$F,$(z-1)$D$)$ for all originator states having $z>0$.

\hspace{5mm}   Let us suppose we are in state $(i$A,$j$F,$z$D$)$ at time $t$ and $l$ of $i$ available nodes have their carrier robot already failed. Note that for classical MDS codes, we need to have $k$ helper nodes to be able to complete the data request successfully\footnote{Various network codes exist that may require to access more than or less than $k$ helper nodes with partial node content accesses for full recovery \cite{Dimakis2011}. The present discussion only slightly changes in case such class of codes are used instead.}. Suppose further that $x$ of these requests are from the failed set, and $k-x$ are from the available and operational ones. Due to sampling without replacement, probability of that happening is given by the hypergeometric distribution\footnote{Sampling with replacement would lead to a Binomially distributed statistics instead.}. In this particular condition, we need to wait for the $x$ failed carriers to be repaired first which is given by the maximum repair time and typically not distributed exponentially. In fact, this distribution can be shown to be equal to the sum of exponential distributions which in this subsection is assumed to be close to another exponential distribution with rate $1/\phi \sum_{m=1}^x 1/m$ where the harmonic sum in the rate can be approximated closely by
\begin{eqnarray}
hs(x) := \sum_{m=1}^x \frac{1}{m} \approx \log(x) + \zeta + \frac{1}{2x} - \frac{1}{12x^2} + \frac{1}{120 x^4} \label{eq_approx}
\end{eqnarray}
where $\zeta = 0.5772156649$ is known as Euler--Mascheroni constant. 

\hspace{5mm}   After all the necessary repair information is collected by any of the $z$ detected nodes, each begins the computation needed for the repair process and write the repaired data to the corresponding storage unit. But the write process needs at least one carrier/robot available. The availability analysis is quite similar to the same case with detection process (each node uses their own robot for detecting the failure) and thus the rate of such happening is represented by $\mu_z(t;\phi)$ expressed as
\begin{eqnarray}
\mu_z(t;\phi) = z\mu - \frac{\mu^2}{\mu + \phi} \sum_{m=1}^{j} (1-\beta_{s_m}(t))
\end{eqnarray}

\hspace{5mm}   On the other hand, the conditional repair rate (conditioned on $x$ and $l$) can be expressed as
\begin{align}
\mu_{iz}(t;\phi,k | x, l) = \frac{\binom{i-l}{k-x}\binom{l}{x}}{\binom{i}{k}} \left( \frac{1}{\mu_z(t;\phi)} + \frac{hs(x)}{\phi}  \right)^{-1}
\end{align}
where $s_m \in \{1,\dots,n\}$. Finally, the unconditional repair rate can be obtained by summing over all $x$ and $l$ as follows,
\begin{align}
    \mu_{iz}(t;\phi,k) &= \sum_{l=0}^i \psi_{i-l}(t) \sum_{x=0}^l \mu_{iz}(t;\phi,k | x, l) \\
    &= \sum_{l=0}^i \psi_{i-l}(t) \mu_z(t;\phi) \\ & + \sum_{l=0}^i \psi_{i-l}(t) \sum_{x=1}^l\mu_{iz}(t:\phi,k | x, l) 
\end{align}

\hspace{5mm}   Note that if $\phi \rightarrow \infty$, i.e., we assume immediate robot repairs, we shall have 
\begin{align}
    \lim_{\phi \rightarrow \infty} \mu_{iz}(t;\phi,k) &= \sum_{l=0}^i \psi_{i-l}(t) \sum_{x=0}^l \frac{\binom{i-l}{k-x}\binom{l}{x}}{\binom{i}{k}} z\mu = z \mu
\end{align}
meaning that robot  repairs  being instantaneous,  the  node repair  rate  would  be $z\mu$ which  is  the same as that of without any robot failures. Finally, we summarize the new state transition table in Table \ref{dest_table3} with indexes satisfying the following inequalities
\begin{align}
k \leq i \leq n,  \\ 
0 \leq j \leq n - k, \\ 
0 \leq z \leq n - k,  \\ 
i + j + z = n. \nonumber
\end{align}

\begin{figure*}[t!]
\centering
  \includegraphics[width=\linewidth]{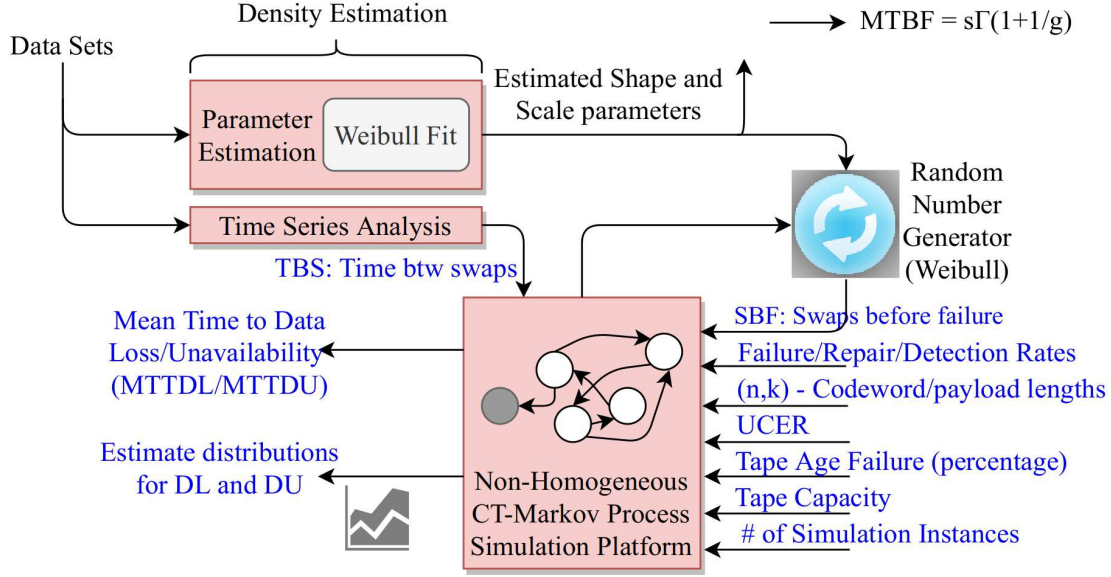}
  \caption{Data-Inspired overall modeling and simulation platform implemented in MATLAB. Density estimation problem is reduced down to parameter estimation through assuming a Weibull distribution for robot exchange performance data. In this figure, $s$ represents the scale and $g$ represents the shape parameters. MSBF: Mean Time Before Failure.}
  \label{fig:modelling}
\end{figure*}

\begin{table}[b!]
\normalsize
\centering
\begin{tabular}{|l|l|l|}
\hline \hline
Current State & Destination State     & Transition Rate \\ \hline
$i$A,$j$F,$z$D    & $(i-1)$A,$(j+1)$F,$z$D & $i\lambda \Delta_i$               \\ \hline
$i$A,$j$F,$z$D    & \textbf{F} & $i\lambda(1-\Delta_i)$             \\ \hline
$i$A,$j$F,$z$D     & $i$A,$(j-1)$F,$(z+1)$D & $\theta_j(t;\phi)$               \\ \hline
$i$A,$j$F,$z$D     & $(i+1)$A,$j$F,$(z-1)$D & $\mu_{iz}(t;\phi,k)$               \\ \hline \hline
\end{tabular}
\caption{State transitions and corresponding transition rates.}
\label{dest_table3}
\end{table}

\subsection{Lower/Upper bounds on the Performance}

\hspace{5mm} For a given finite carrier repair rate $\phi < \infty $, if we let the exchange rate tend to large values the carrier repairs will not be able to catch up,  eventually resulting in total carrier unavailability. In that particular case, it is of interest to drive the lower bound on performance in a closed form. We note that in case of total carrier unavailability,  there is no failure detection and therefore it means no data repair in a cold storage context and hence, the survival time depends on which state the system is in and whether the hard error leads to unrecoverable state transitions. In light of this observation, the lower bound ($LB$) can be derived.

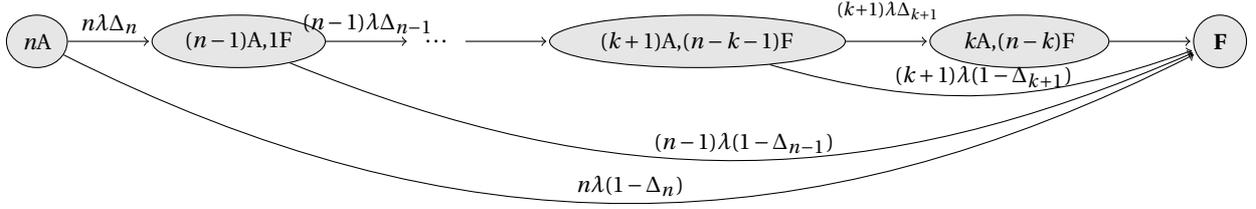
\begin{figure}[t!]
\centerline{
\begin{tikzpicture}[node distance=1.1cm]
    %\hspace*{-1.5cm}
    \tikzset{node style/.style={state, fill=gray!20!white}}
    \tikzset{node style2/.style={state, fill=gray!80!white}}
    % Add the states
    \footnotesize
    \node[node style, ellipse] (n)   {$n$A};
    \node[node style, ellipse, right=of n] (n1) {$(n-1)$A,$1$F};
    \node[state,  line width=0.4mm, right=of n1, draw=none] (n2) {$\dots$};
    \node[node style, ellipse, right=of n2] (j) {$(k+1)$A,$(n-k-1)$F};
    \node[node style, ellipse, right=of j] (m1) {$k$A,$(n-k)$F};
    \node[node style, ellipse,  right=of m1] (m) {$\textbf{F}$};

    % Connect the states with arrows
    \draw[every loop]
        (n) edge[bend left=0, auto=left] node {$n\lambda \Delta_n$} (n1)
        (n) edge[bend right=27, auto=left] node {$n\lambda (1-
        \Delta_n)$} (m)
        (n1) edge[bend right=23, auto=left]  node {$(n-1)\lambda (1-\Delta_{n-1})$} (m)
        (n1) edge[auto=left] node {$(n-1)\lambda \Delta_{n-1}$} (n2)
        (n2) edge[auto=left] node {} (j)
        (j) edge[auto=left] node {$(k+1)\lambda \Delta_{k+1}  \atop$} (m1)
        (m1) edge[auto=left] node {} (m)
		(j) edge[bend right=18, auto=left]  node {$(k+1)\lambda (1-\Delta_{k+1})$} (m);
\end{tikzpicture}
}
\caption{The general CT Markov model reduces to single dimensional one with the following transition rates.}\label{fig:markov2422}
\end{figure}

\hspace{5mm} Let us start with rewriting the conditional probability that we do not end up in the total data loss (failure) state when there are $i$ available nodes, $\Delta_i$ in an explicit integral form
\begin{eqnarray}
\Delta_i &=& \frac{\Gamma(i+1)}{\Gamma(i-k)\Gamma(k+1)} \int_{\eta(t)}^1 v^{a-1} (1-v)^{b-1}dv \nonumber \\
&=& \frac{i!}{(i-k-1)!k!} \int_{\eta}^1 v^{i-k-1} (1-v)^{k}dv \nonumber \\
&=& (i-k) \binom{i}{k} \int_{\epsilon+\kappa - \epsilon\kappa}^1 v^{i-k-1} (1-v)^{k}dv
\end{eqnarray}
where $\eta = 1 - (1-\epsilon)(1-\kappa)$ is the hard error rate. To derive the lower bound we consider the case of total carrier unavailability. Thus, this results in no failure detection and henceforth no data repair process is initiated. This leads to the simplified version of the Markov model as shown in Fig. \ref{fig:markov2422}. Note that the hard error that leads to total failure when there are $i$ available nodes happens with probability $(1-\Delta_i)\prod_{j=i+1}^n \Delta_j$. The total time before failure is the sum of average times spent in each of the visited system states i.e., $1/n\lambda, 1/(n-1)\lambda, \dots, 1/i\lambda$. Finally, by summing over all possible $i$, we can estimate the lower bound on the total average time spent before failure as
\begin{eqnarray}
LB &=& \sum_{i=k}^n (1 - \Delta_i) \prod_{j=i+1}^n \Delta_j \sum_{j=i}^n \frac{1}{j\lambda}  \nonumber \\
&=& \sum_{i=k}^n (1 - \Delta_i) \prod_{j=i+1}^n \Delta_j \left( hs(n) - hs(i-1) \right) \nonumber \\
&\approx& \sum_{i=k}^n (1 - \Delta_i) \prod_{j=i+1}^n \Delta_j  \Bigg( \log\left(\frac{n}{i-1}\right)^{1/\lambda}    + \frac{1}{\lambda}\sum_{l=0}^1(-1)^l\left(\frac{1}{2n_l} - \frac{1}{12n_l^2} + \frac{1}{120n_l^4}\right) \Bigg) \label{lowerbound}
\end{eqnarray}
with $\Delta_k = 0$ and $n_l = n - l(n-i+1)$. Note  that if there is no hard errors, i.e., $\Delta_i = 1$ for $i=n,\dots,k+1$, we shall have the simplified lower bound with the closed form expression given by $\sum_{j=k}^n 1/j \lambda$. In general we have
\begin{eqnarray}
LB \leq \sum_{j=k}^n \frac{1}{j \lambda} =hs(n) - hs(k-1) \label{lowerbound}
\end{eqnarray}
Note that the approximation follows due to  \eqref{eq_approx}.

%is approximated in \ref{appendixB} in terms of %$\Delta_i$'s as follows
%\begin{eqnarray}
%LB &\approx \sum_{i=k}^n (1 - \Delta_i) \prod_{j=i+1}^n \Delta_j  \Bigg( \log\left(\frac{n}{i-1}\right)^{1/\lambda}  \nonumber + \frac{1}{\lambda}\sum_{l=0}^1(-1)^l\left(\frac{1}{2n_l} - \frac{1}{12n_l^2} + \frac{1}{120n_l^4}\right) \Bigg) \label{lowerbound}
%\end{eqnarray}
%with $\Delta_k = 0$ and $n_l = n - l(n-i+1)$. 

\hspace{5mm} On the other hand, if we let the exchange rate tend to  zero there will be no need for  carrier  repairs,   resulting  in  total carrier availability. In that particular case, it is of interest to drive the  upper bound on performance  in  a  closed form. We realize that there is only one absorbing state in our model (Failure state) and hence, the TPM is already in its canonical form,
\[ \textbf{P}_{N_s \times N_s} = \left[
    \begin{array}{c;{3pt/3pt}r}
    \mbox{$\textbf{L}_{N_s-1 \times N_s-1}$} & \mbox{ $\textbf{R}_{N_s-1 \times 1}$} %\begin{matrix} 0 \\ 0 %\end{matrix} 
    \\ \hdashline[3pt/3pt]
    \begin{matrix} 0 & \dots & 0 \end{matrix} &  1
    \end{array}
    \right]
\]

\hspace{5mm} For an absorbing Markov chain, we know that the inverse of $\textbf{I} - \textbf{L}$ matrix is called the fundamental matrix (denoted as $\textbf{M}$) and it can be expressed as
\begin{eqnarray}
\textbf{M} = (\textbf{I}-\textbf{L})^{-1} = \textbf{I} + \sum_{i=1}^{\infty} \textbf{L}^i
\end{eqnarray}
in which $m_{ij}$ entry provides the expected number of times that the Markov process visits the transient state $s_j$
when it is initialized in the transient state $s_i$. Since we initially assume all nodes to be available in the beginning, we are interested in $m_{1j}$s i.e., the system is assumed to be in state $n$A in the beginning of the operation. Since for $s_j$, all outgoing transitions happen according to exponential distributions and the hold time is given by the minimum which  is also distributed exponentially with rate $-q_{jj}$. This implies the average hold time in each visit to $s_j$ is given by $-1/q_{jj}$. Finally, the upper bound can be approximated by
\begin{eqnarray}
UB \approx -\sum_{j=1}^{N_s-1} \frac{m_{1j}}{q_{jj}} \label{upperbound}
\end{eqnarray}

\hspace{5mm} Note that this is only an approximation since TPM is an approximation to the continuous time Markov model. Also, we can analytically assess the upper bound on the time-dependent performance including the robot failure and repair processes by considering only the rate matrix given by Table \ref{dest_table3} instead of Table \ref{dest_table2}. This is possible because we have approximated distributions as exponential to keep Markovianity intact. This approximation will later in numerical results section be verified to be sufficiently accurate for the range of parameters of interest.

\subsection{A Data-Assisted Modeling Framework}

\hspace{5mm} As the internal engineering details of storage devices get more complicated, data-centric reliablity analysis gains attraction. In this section, we present an instantiation of such an effort for cold/tape storage scenario. In this end, we utilize a data-inspired approach for estimating the number of round-trips (exchanges in our context) that a carrier make before a critical failure happens. The critical failure takes place when the robot is no longer able to operate within the library system due to various reasons till they are replaced with the new one. In our tape application, the total number of robot exchanges before failure (SBF) is assumed to be Weibull distributed which shall be validated by the collected field data using enterprise Quantum libraries. Weibull distribution is completely characterized by two independent parameters called the shape ($g$) and scale ($y$). The reason we choose Weibull is  twofold. First, it is the generalization of the most commonly assumed exponential distribution (single parameter) in literature. In other words, by selecting appropriate parameter values Weibull can be transformed to exponential distribution. Secondly, it is heavy tailed and closely characterize the observed field data. We realize that the heavy-tailed distributions characterize various types of data accurately as the number parameters of the distribution increase. For instance, it is reported in various studies  that the data object size tends to possess heavy-tailed distribution such as Pareto \cite{Satyanarayanan}. On the other hand, several studies show that heavy-tail distributions might well characterize local file system dynamics and file sizes \cite{downey2001structural}, archival data \cite{ramaswami2014modeling} and the data stored and communicated over the world wide web \cite{gong2001tails}. 

\hspace{5mm} We note that based on the available field data and Weibull assumption, the challenging density estimation problem is transformed into parameter estimation problem. More precisely, we estimate the shape ($g$) and scale ($y$) parameters of the distribution through simple linear regression. Secondly, we obtain an estimate of the distribution of the time between exchanges/swaps. Using the same data set, this distribution is observed to have exponential tail and hence a single parameter (the rate) will have to be estimated. An exponential assumption is also quite nifty because the corresponding count process will become analytically tractable Poisson distribution. Since the estimated parameter is a function of the utilization rate of the system and hence is time-dependent, we shall test a range of values in our simulations to illustrate the overall picture. A summary of the modeling framework is depicted in Fig. \ref{fig:modelling}. In this framework, the data-based parameter estimations ($\hat{g}$ and $\hat{y}$) are fed into the proposed non-homogeneous Markov Process as estimated inputs.  In addition to these inputs, we also set the rest of the simulation parameters $\lambda,\mu,\theta,n,k,\kappa,\epsilon$ as well as the number of simulation instances to some appropriate values based on the field data and our experience with 6TB tapes. The system is protected with a $(n,k)$ MDS code. The random number generator chooses a random SBF value according to the estimated Weibull distribution and repeats this process and uses a unique realization at each iteration of the simulation. We typically simulate over 10000 times to obtain reliable values.

\hspace{5mm} The main purpose of the simulation platform is to estimate the distribution of the overall data loss and/or unavailability (which ever one  degrades the performance first) at the same time to demonstrate the implicit relationship of these two important performance metrics. In other words, we can finally numerically estimate MTTDL and data MTTDU metrics quite confidently. In addition, the mean value of the number of exchanges is given by $\hat{y}\Gamma(1 + 1/\hat{g})$ which shall be used as the guideline of robot performance in the numerical results section.

\begin{figure}[t!]
\centering
  \includegraphics[width=0.8\linewidth]{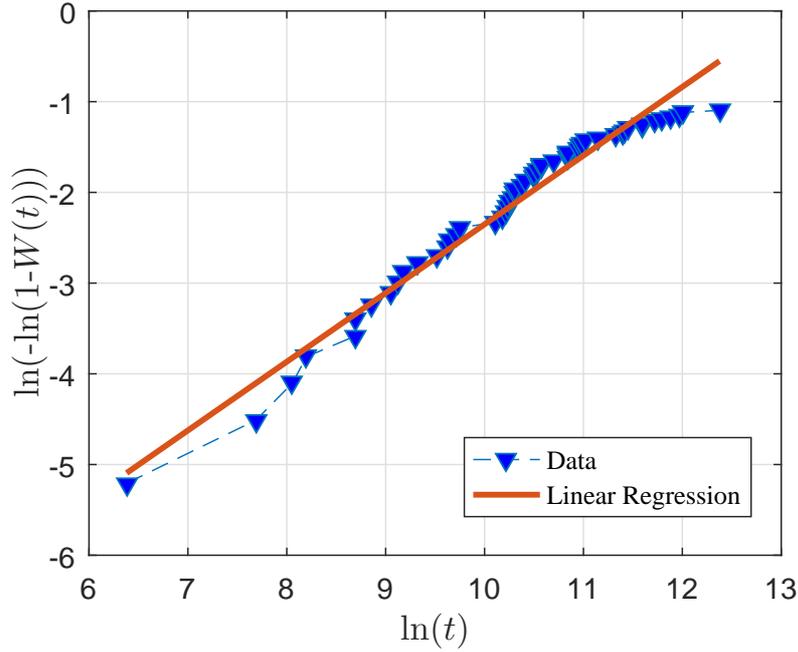}
  \caption{Robot exchange data and Weibull parameter estimations using a linear regression.}
  \label{fig:data_robot_xph}
\end{figure}

\hspace{5mm} Note that there are more than one way for the estimation of the Weibull distribution parameters i.e., $g$ and $y$. We adapt simple linear regression in this study to estimate these parameters of the Weibull distribution. However, few algebraic manipulations are needed to put the CDF of the Weibull in an appropriate form. Accordingly, let us remember the Weibull CDF $W(t)$ as given by the equation.
\begin{eqnarray}
W(t) = 1 - e^{(t/y)^g}
\end{eqnarray}
which can be rearranged and expressed as the following linear equation
\begin{eqnarray} \ln\left(-\ln\left(1-W(t)\right)\right) =  g\ln(t/y) =  g\ln(t) - g\ln(y) \label{eqn411}
\end{eqnarray}
If we set the ordinate to the left hand side, $\ln\left(-\ln\left(1-W(t)\right)\right)$ and abscissa to $\ln(t)$, and apply a linear regression, we shall have a linear function that will naturally have an intercept($\mathcal{I}$) and a slope($\mathcal{S}$). Using these estimates we can generate the estimates of the shape parameter ($\hat{g}$) as well as the scale parameter ($\hat{y}$) as  shown below,
\begin{eqnarray}
\hat{g} = \mathcal{S}, \ \ \hat{y} = \exp{\left(-\mathcal{I}/\hat{g}\right)}. \label{eqn422} 
\end{eqnarray}

\hspace{5mm} In other words, the slope of the line shall be the shape parameter whereas the scale parameter needs to be calculated based on the estimate of the shape parameter according to equation \eqref{eqn422}. To demonstrate the accuracy of the Weibull assumption, we recorded around 40000 robot exchanges before they cease operation. These equal-quality robots are operating inside Quantum Scalar i6K enterprise libraries which can house up to 12000 cartridges and is optimized for high density. This data is plotted in Fig. \ref{fig:data_robot_xph} based on the formulation given in equation \eqref{eqn411}, where the intercept and slope can easily be found and used to calculate the shape parameter, $\hat{g}=0.76$ and scale parameter $\hat{y}=491669$. The accumulation in the data for $t$ values satisfying $9 \leq \ln(t) \leq 12$ is due to the fact that most robots have a logarithmic lifetime in that range. Based on the estimated parameters of the Weibull distribution, the average number of exchanges can be calculated to be $\hat{y}\Gamma(1 + 1/\hat{g}) = 580747$ exchanges before critical robot failure happens. In the numerical results section, we shall choose our parameters within the ballpark of these figures to make our results/conclusions realistic. We finally note that the use of distributions with more parameters could approximate the data better, however  it will only result in extremely minor accuracy advantage at the expense of increased estimation complexity.

\subsection{Numerical Results}

\hspace{5mm} As it is usually the case with cold (and archival) storage platforms, we primarily focus on the read-back or in other words the data retrieval performance. Few numerical results for the proposed simulation and modeling platform are presented. The intention is to illustrate reliability (in terms of MTTDL) and unavailability (in terms of MTTDU) in the same plot on the ordinate as a function of other simulation parameters. The abscissa  could be one of the parameters of the system including the exchange and carrier repair rates. Since the number of simulation parameters are plenty and it is hard to visualize/plot higher dimensional data using two dimensions, we present a 2-D plot where we fix most of the simulation parameters except the exchange/swap (\textit{xph}) and carrier repair rates ($\phi$). The former typically changes based on the system utilization rate whereas the latter is under the control of system maintenance team. Another reason for choosing these parameters to vary is that they directly affect the unavailability of the system i.e., in the absence of the carrier (failed carrier) overall data access time increases until carrier repair takes over. We particularly note that most of the carrier devices (e.g. robots) are shipped with a maximum exchange/swap rate number for reliable operation (such as 840 exchanges per hour (\textit{xph}) \cite{richards2018}), we vary the abscissa from some small exchange number to somewhere above the reported maximums and present results in a log-log plot. Similarly, when we plot the MTTDU in terms of $\phi$, we fixed the exchange/swap rate and varied the carrier repair rate to see the effect of repair frequency on the unavailability performance. Please note that the system could be operating at any point on these performance curves at a given time $t$.

\begin{table}
    \normalsize
    \caption{Simulation parameters}
    \begin{tabularx}{\columnwidth}{X|X}
        \hline
        Parameter                 & Value    \\
        \hline
        $\lambda$ (hours) & 1/50000 \\
        $\mu$ (hours)    & 1/24       \\
        $\theta$ (hours)       & 1/8760 \\
        $(n,k)$         & Variable \\
        $\epsilon$ (UCER)    & $10^{-19}$      \\
        $tape\_capacity$ & 6TB \\
        $\hat{g}$ (shape, Weibull) & 0.37 and 0.67 \\
        $\hat{y}$ (scale, Weibull) & 525985 \\
        $\kappa$ & 0.001 \\
        $\#$ of simulations & $>$10000 \\
        \hline
    \end{tabularx}
    \label{table:simulation parameters}
\end{table}

\begin{figure}[t!]
\centering
  \includegraphics[width=0.7\linewidth]{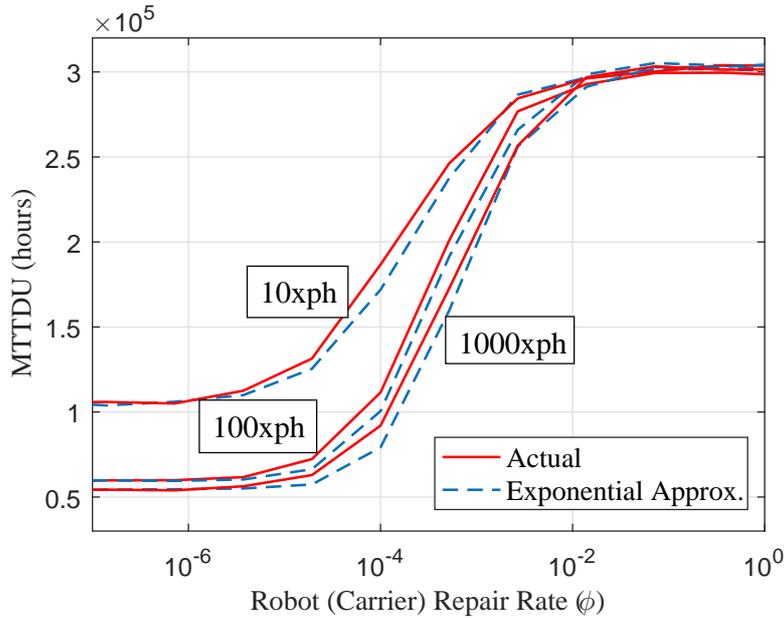}
  \caption{The accuracy of exponential-tail approximation with respect to MTTDU as a function of robot repair rate ($\phi$) using a (4,2) MDS code for two different exchange rates 10xph and 100xph.}
  \label{fig:modelling_comp}
\end{figure}

\begin{figure}[t!]
\centering
  \includegraphics[width=0.7\linewidth]{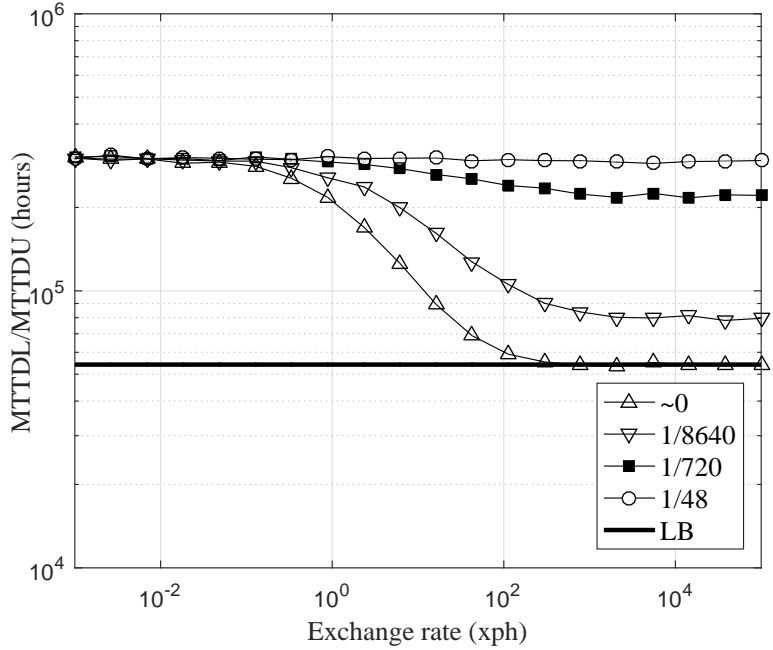}
  \caption{MTTDL/MTTDU (hours) as a function of Exchange rate (round trips or swaps per hour) and a (4,2) MDS codes with the shape parameter 0.67 for different repair rates $\phi$. LB stands for the derived lower bound.}
  \label{fig:modelling1}
\end{figure}

\hspace{5mm} The parameters of the simulation are briefly summarized in Table \ref{table:simulation parameters}. As can be seen, we have assumed a day-long mean data/tape repair and a year-long mean failure detection time as a starting point. These numbers again are application specific and can be changed per use case. We have selected few example half-rate code parameters such as $(4,2)$ and $(6,3)$ with varying reliability guarantees. The scale--out system consists of $n$ identical libraries where each library stores and handles only one chunk of data when requested. Each library has their own unique robot and contains multiple storage units such as tapes. Tapes and robots are assumed to be of equal quality and type. The direct effect of inner details of the scale-out library system such as the total number of libraries $M$, the number of tapes per library, the geometry of the tape shelf locations etc. are accounted by the Weibull  parameter estimations (shape and scale) and data-assisted modeling framework. This data analysis saves us from getting into the inner complexities of library systems and provides us the statistical nature of the number of exchanges per library. This is later used as an input for the proposed generalized Markov model introduced in the previous section. Finally, note that UCER is assumed to be $10^{-19}$ which is way lower than $10^{-15}$, the UCER of the known disk drive systems. This is due to the high data durability guarantees of the next generation tape technology \cite{dholakia2008new}.

\hspace{5mm} Our first simulation is presented in Fig. \ref{fig:modelling_comp} where we clearly demonstrate the validity of our exponential tail assumption made earlier (such as the expression \eqref{exptailapprox1}) for approximating the non-exponential distributions that appear in various stages of the proposed Markov model. We have used (4,2) MDS code and three exchange rates namely 10xph, 100xph and 1000xph and plotted MTTDU results using both the actual distributions as well as the exponential-tail approximation (used in this simulation) as a function of robot repair rate. One of the initial observations is that our exponential-tail assumption leads to a lower bound on the actual MTTDU values. Furthermore, the worst case difference between the actual and approximate MTTDU values do not affect the \textit{number of nines}, a metric typically used to express system reliability with regard to MTTDL performance metric in the industry. For the rest of this subsection, we shall present our results using the exponential-tail approximation due to simpler formulation as well as analytical tractability. 

\hspace{5mm} In Fig. \ref{fig:modelling1}, we present MTTDL/MTTDU in hours as a function of exchange rate for a (4,2) MDS code. In light of our data observations and equations \eqref{eqn411} and \eqref{eqn422} derived earlier for estimating the scale and shape parameters, we have found that two shape parameters $0.37$ and $0.67$ along with the scale parameter $525985$ are most common giving us the average total number exchanges of  $695563$ and $2200634$, respectively, before a robot failure happens. Note that if a robot lasts only after one year, these numbers would indicate an average of 79.4 xph and 251.2 xph, respectively. We have also included the availability lower bound (as given by the equation \eqref{lowerbound}) in our plots which do not change with the growing exchange rate. 

\hspace{5mm} As can be seen from Fig. \ref{fig:modelling1}, as the exchange rate tends to zero (going from right to left on abscissa), the reliability closely converges to the durability of the system model introduced in Fig. \ref{GMMs3} where the availability issue posses no more risk to data access anymore. On the other hand, as the exchange rate tends to large values (going from left to right on abscissa), MTTDU converges to the durability lower bound. Depending on the operating point of the library system, our model clearly shows how the unavailability changes as a function of exchange rate if we do not have sufficiently frequent robot repair in place. Also, it can be observed from the same simulation data that a robot repair rate of $\phi = 1/48$ seems to be sufficiently frequent for the system maintenance and hence we do not see any notable  reduction in the availability for this particular  repair rate.

\begin{figure}[t!]
\centering
  \includegraphics[width=0.7\linewidth]{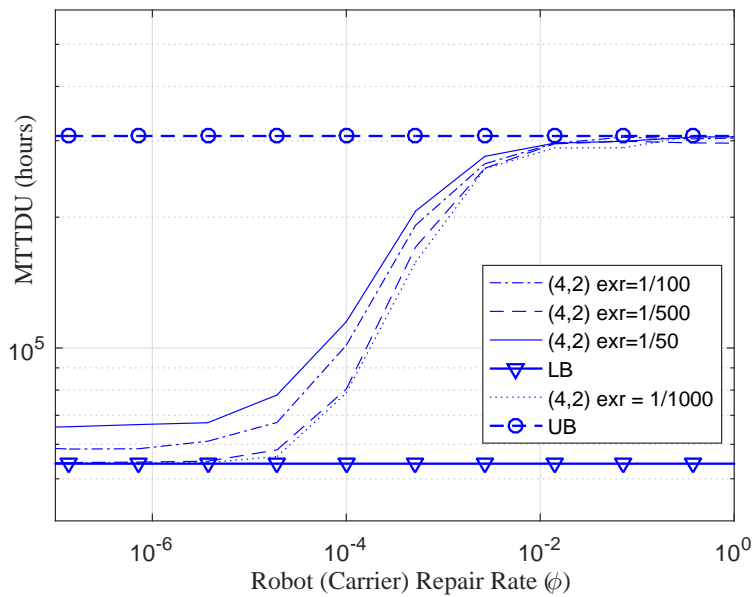}
  \caption{MTTDU (hours) as a function of robot repair rate ($\phi$) and few example exchange rates (exr) shown for (4,2) MDS code. LB stands for the lower bound.}
  \label{fig:modelling_phi1}
\end{figure}

\begin{figure}[t!]
\centering
  \includegraphics[width=0.7\linewidth]{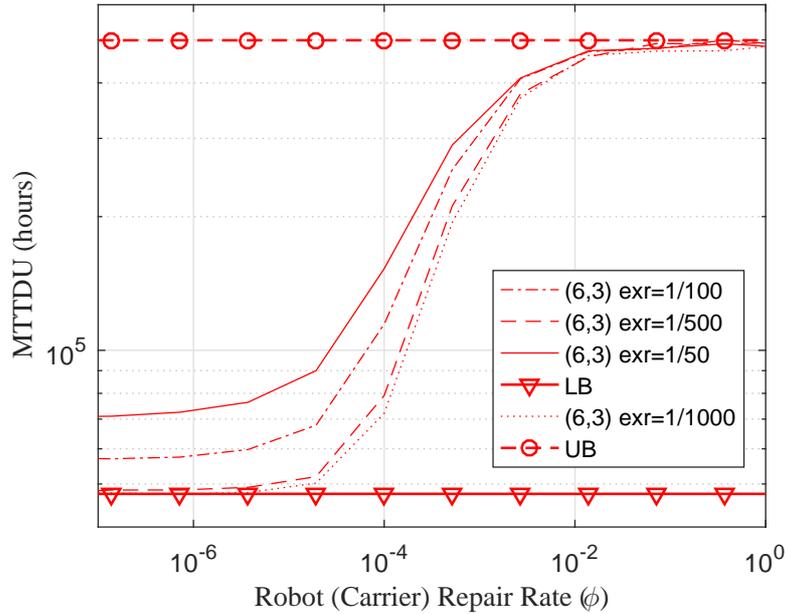}
  \caption{MTTDU (hours) as a function of robot repair rate ($\phi$) and few example exchange rates (exr) shown for (6,3) MDS code. LB stands for the lower bound.}
  \label{fig:modelling_phi2}
\end{figure}

\hspace{5mm}  On the other hand, we observe from Fig. \ref{fig:modelling1} (and for that matter in Figs. \ref{fig:modelling_phi1} and  \ref{fig:modelling_phi2}) that as we increase the robot repair rate $\phi$, i.e., we perform more frequent robot repairs, we improve the availability. However, at some point, increasing $\phi$ does not help us much i.e., system's robots are repaired fast enough that no unavailability leads to a dramatic performance loss.  In order to find the optimal robot repair rate, we also need to plot MTTDU as a function of $\phi$ for a range of exchange/swap rates. These performance plots are shown in Fig. \ref{fig:modelling_phi1} and Fig. \ref{fig:modelling_phi2} for the half-rate MDS codes (4,2) and (6,3), respectively. Also included in the same plots are the corresponding lower and upper bounds computed using equations \eqref{lowerbound} and \eqref{upperbound}, respectively. There are two interesting observations common to both plots. One of the observations is that MTTDU performances converge after certain exchange rates. For example in Fig. \ref{fig:modelling_phi1}, $\phi=1/500$ and $\phi=1/1000$ do not provide dramatically different MTTDU performances. Exact same trend can be observed for (6,3) MDS code in Fig. \ref{fig:modelling_phi2} as well. Therefore, since lowering the exchange rate improves the availability, we can talk about an optimal repair rate beyond which we do not experience any unavailability for all possible exchange rates of interest. Having determined the optimal exchange rate for a given system is important from both user satisfaction and energy savings point of views. The second observation with respect to these plots is that by keeping the code rate fixed, as the blocklength of the MDS code gets larger, the associated lower bound gets worse. This is due to the number of parities do not scale as much as needed to compensate for the increased blocklength. However, using the expressions derived for the lower bound (the equation \eqref{lowerbound}), one can immediately notice that the performance difference between different size MDS codes of the same rate will disappear as $n$ tends large. From Fig. \ref{fig:modelling_phi1} and Fig. \ref{fig:modelling_phi2}, we can quantify this difference for both half-rate codes, namely (4,2) and (6,3) MDS codes, respectively.

\section{Advanced Modeling}

\hspace{5mm} In our previous considerations, the set of extensions and Markov analysis associated with these models can be argued to capture a simplistic view of an actual storage system, particularly while the size and the scale of such systems are ever growing. Although, the previous discussions are quite improved versions of the available analytical methods using real data (this makes them great machineries for back-of-the-envelope comparisons), they can still be criticized due to the underlying model assumptions such as constant failure and repair rates. Unfortunately though, the inclusion of time dependence into reliability models increases complexity
and can prohibit analytic reliability estimates. In addition, we have observed that extending the canonical model to multi-disk fault-tolerant systems leads to issues in
modeling rebuild due to memoryless property of underlying exponential failure/repair time distribution.

\subsection{Incorporating sector errors}

\hspace{5mm} Previous models given for incorporating hard errors into the Markov model suffers from representing accurately the latent sector errors which are found to be more frequent than unrecoverable hard bit read errors \cite{Bair}. In the literature, there have been two ways to model the latent sector failure/scrubs. One is to treat them as disk failures or model them as a separate random processes. Here we present the latter approach which is based on regular scrubbing to remedy the latent sector errors. Given a fixed scrub period for a given sector,
denoted as $T_S$, load on a given sector, $l$, probability of a sector error due to a write operation,
$P_w$ and the ratio of write requests in the total system load, $r_w$, the probability of an
unrecoverable error on a given sector at an arbitrary time $t$ (not time dependent) is given by \cite{Ilias}
\begin{eqnarray}
P_S = \left(1 - \frac{1 - e^{-lT_S}}{lT_S}\right)P_wr_w
\end{eqnarray}

In the same study, a time-dependent probability is derived to be of the form
\begin{eqnarray}
P_S(t)  = \left(1 - e^{-l(t \mod T_S)}\right) P_e
\end{eqnarray}
where $P_e$ is the single sector failure probability. Unfortunately, incorporating the load term $l$, into such a calculation to come up with a proper sector failure model remains an open problem. $P_S$ can be used to calculate the likelihood of a latent sector failure on a disk.
If a disk has $S$ sectors, then the probability that one or more latent sector errors are
present at an arbitrary point in time is $P_{LSE} = 1 - (1 - P_S)^S$. Given the fact that sector errors are more dominant type of failures relative to unrecoverable bit errors, $P_{LSE}$ can be used in place of $P_{UCER}$ given in the Markov model that incorporates the hard error in the MTTDL calculations (See equation \ref{PUCER}). Although such a change shall create more accurate estimates, it does
not account for the critically exposed region of the rebuild process or the number of
sector errors \cite{GreenanThesis}.

\subsection{Simulators}

\hspace{5mm} With all the extensions of the canonical Markov model, the remaining critical problem with previously discussed Markov models is the time dependent failure and repair rates. In addition, they are argued not to model well multi-disk fault tolerance, irregular fault-tolerance and sector errors. Due to these reasons, the most effective way remaining is simulation. Standard simulation tools handles time dependence pretty effectively, however obtaining a result in a reasonable amount of time is non-trivial (mostly due to rare events). Standard simulation tools use two methodologies to evaluate the reliability of an erasure coded system \cite{Nicola}.  One of them is based on the estimation of unreliability function $U(t)$ and the other is based on MTTDL. Let $T_F$ be the random variable characterizing the failure time, $T_i$ be the stopping time of the $i$th iteration and $\textbf{1}_{T_F \leq t}(t)$ be the indicator function, using a law of large numbers argument we have
\begin{eqnarray}
\hat{U}(t) &=& \frac{1}{N} \sum_{i=1}^N \textbf{1}_{T_F \leq t}(t) \\
MTTDL &=& \frac{1}{N} \sum_{i=1}^N T_i
\end{eqnarray}
where the simulation is assumed to carry out $N$ consecutive iterations. As previously noted, aforementioned simulation methods
are inefficient when evaluating highly fault tolerant systems. The major drawback to this type of simulation is the amount of time required to
get an accurate result when the probability of failure is extremely low. The main approach to circumvent this situation is to increase the frequency of rare events after some threshold number of failures occur in the standard simulation tool. This technique is generally called \emph{Importance Sampling} in literature \cite{Marvin}. Main objective is to increase the probability of seeing data loss event while keeping the variance (system) reduced after the threshold is reached.

\hspace{5mm}  There have been many more advanced simulation-based reliability estimators, renowned by their high fidelity book keeping (tracking which disk and sector failures have occurred, and efficiently determining if the failures constitute a data loss event) features. One of them is High-Fidelity Reliability (HFR) Simulator. For more information and the validation routines, the reader is referred to the reference \cite{GreenanThesis}.

%----------------------------------------------------------------------------------------
%	Section 2
%----------------------------------------------------------------------------------------

%\section{Linear Codes}

%------------------------------------------------

%----------------------------------------------------------------------------------------

\end{document}